\documentclass[twoside]{IEEEtran}
\usepackage{amsmath,amsfonts}
\usepackage{algorithmic}
\usepackage{algorithm}
\usepackage{array}
\usepackage[caption=false,font=normalsize,labelfont=scriptsize,textfont=scriptsize]{subfig}
\usepackage{textcomp}
\usepackage{url}
\usepackage{verbatim}
\usepackage{graphicx}
\usepackage[noadjust]{cite}
\usepackage{multirow}
\usepackage{epstopdf}

\usepackage{flushend}
\usepackage[T1]{fontenc}
\usepackage{balance}

\begin{document}
\title{Convex Hull Prediction for Adaptive Video Streaming by Recurrent Learning}

\author{Somdyuti Paul,
        Andrey Norkin,
        and Alan C. Bovik
\thanks{S. Paul and A. C. Bovik are with the Department
of Electrical and Computer Engineering, University of Texas at Austin, Austin,
TX, 78712 USA (email: somdyuti@utexas.edu, bovik@ece.utexas.edu).}
\thanks{A. Norkin is with Netflix Inc. Los Gatos, CA, 95032 USA (email: anorkin@netflix.com).}
\thanks{This work is supported by Netflix Inc.}}



\markboth{IEEE Transactions on Image Processing}%
{Paul \MakeLowercase{\textit{et al.}}: Convex Hull Prediction for Adaptive Video Streaming by Recurrent Learning}

\IEEEpubid{\begin{minipage}{\textwidth}\ \\[12pt] \\
\copyright 2024 IEEE. Personal use of this material is permitted. Permission from IEEE must be obtained for all other uses, in any current or future media, including reprinting/republishing this material for advertising or promotional purposes, creating new collective works, for resale or redistribution to servers or lists, or reuse of any copyrighted component of this work in other works.\\ 
\end{minipage}} 

\maketitle

\begin{abstract}
Adaptive video streaming relies on the construction of efficient bitrate ladders to deliver the best possible visual quality to viewers under  bandwidth constraints. The traditional method of content dependent bitrate ladder selection requires a video shot to be pre-encoded with multiple encoding parameters to find the optimal operating points given by the convex hull of the resulting rate-quality curves. However, this pre-encoding step is equivalent to an exhaustive search process over the space of possible encoding parameters, which causes significant overhead in terms of both computation and time expenditure. To reduce this overhead, we propose a deep learning based method of content aware convex hull prediction. We employ a recurrent convolutional network (RCN) to implicitly analyze the spatiotemporal complexity of video shots in order to predict their convex hulls. A two-step transfer learning scheme is adopted to train our proposed RCN-Hull model, which ensures sufficient content diversity to analyze scene complexity, while also making it possible to capture the scene statistics of pristine source videos. Our experimental results reveal that our proposed model yields better approximations of the optimal convex hulls, and offers competitive time savings as compared to existing approaches. On average, the pre-encoding time was reduced by 53.8\% by our method, while the average Bj{\o}ntegaard delta bitrate (BD-rate) of the predicted convex hulls against ground truth was 0.26\%, and the mean absolute deviation of the BD-rate distribution was 0.57\%.

\end{abstract}

\begin{IEEEkeywords}
Per shot encoding, convex hull, bitrate ladder, adaptive streaming, Conv-GRU, recurrent convolutional network.
\end{IEEEkeywords}

\section{Introduction}
\label{sec:intro}
\IEEEPARstart{T}{he} past decade has witnessed a rapid increase in consumer video consumption over the internet, with video-on-demand (VOD) traffic currently constituting over  70\% of all mobile data traffic, and projected to increase to over 80\% by the end of 2028 \cite{ericsson}, outpacing the growth of network bandwidths over the same period. Adaptive streaming solutions are widely employed by streaming service providers to deliver videos having the highest possible visual quality, while also meeting constraints imposed by network bandwidth limitations and other factors such as client device resolutions. This process requires that multiple versions of each video content is encoded, each using different encode configurations, allowing client applications to select and stream a suitable version of the encoded content based on bandwidth and resolution constraints at their end. The decoded videos can be upsampled to the device resolution of the client whenever a lower resolution is streamed. HTTP adaptive streaming (HAS) protocols, such as the HTTP live streaming (HLS) protocol \cite{apple} developed by Apple for H.264 and H.265, and the codec agnostic, Dynamic Adaptive Streaming over HTTP (DASH) \cite{dash}, that was standardized by MPEG, have received wide adoption in adaptive streaming over the internet. However, these HAS protocols do not specify the bitrate adaptation logic, allowing streaming servers the flexibility to implement different versions of bitrate adaptation to support different use cases. Alternative mechanisms for varying the bitrate and quality of video stream have also been developed. Scalable video coding is one such mechanism, which encodes videos at multiple layers to support enhancements of frame rate, picture size, picture quality, etc. A similar mechanism is provided by multiple description coding, which allows multiple encoded descriptions to be transmitted, and each description to be decoded independently, with quality being proportional to the number of descriptions received and decoded. However, these mechanisms of bitrate adaptation often cause losses of compression efficiency due to associated overheads, and produce more complex codecs. Unlike these schemes, adaptive bitrate streaming can support switching between different video coding formats, and is also more suitable for incorporating legacy decoders.  All these factors have contributed to its more widespread adoption.

\IEEEpubidadjcol The earliest adaptive streaming solutions were implemented using fixed, content agnostic encoding recipes, which produced static bitrate ladders as instantiated in \cite{pertitle}, where predetermined pairs of bitrates and resolutions for all contents are specified, irrespective of their genre or visual complexity. However, streaming contents vary widely in texture, motion, and scene complexity, which means that a given bitrate does not correspond to the same level of perceptual quality across contents. Because of this, any content agnostic approach is sub-optimal. Consequently, the paradigm of using static bitrate ladders for adaptive streaming was preempted by the introduction of per-title encode optimization methods \cite{pertitle}, whereby encoding complexity analysis is carried out independently on each title in the video catalog by pre-encoding the sources at different resolutions and bitrates. This is followed by identifying the boundary of the convex hull of the resulting set of bitrate-quality points. The convex hull, as mathematically defined in Section \ref{subsec:convex_hull}, completely describes the bitrate ladder of the corresponding video content. Each content is streamed according to this customized ladder. The adaptive convex hull approach was further refined using per-shot encoding \cite{pershot}, whereby each video is segmented into its constituent shots before performing pre-encoding analysis to determine the bitrate ladder for each shot. This content aware, per-shot framework leads to a substantial reduction in the Bj{\o}ntegaard delta bitrate (BD-rate) \cite{bjontegaard} over the best fixed quantization parameter (QP) encoding scheme, when using modern video codecs such as VP9 \cite{vp9} and HEVC \cite{hevc}, in terms of both traditional peak signal to noise ratio (PSNR), as well as the perceptually-motivated video multimethod quality assessment (VMAF) \cite{vmaf} quality model \cite{do}. Nevertheless, this increased encode efficiency comes at the cost of significantly increased computation, since a large number of pre-encodes must be generated corresponding to the different encoding parameters, even though only a small fraction of all the different encoded versions usually lie on the convex hull and are eventually streamed. Thus, the process of generating convex hulls per video shot for adaptive streaming is both a time consuming and a resource intensive process. 

\begin{figure*}
\centering
    \includegraphics[width=18cm]{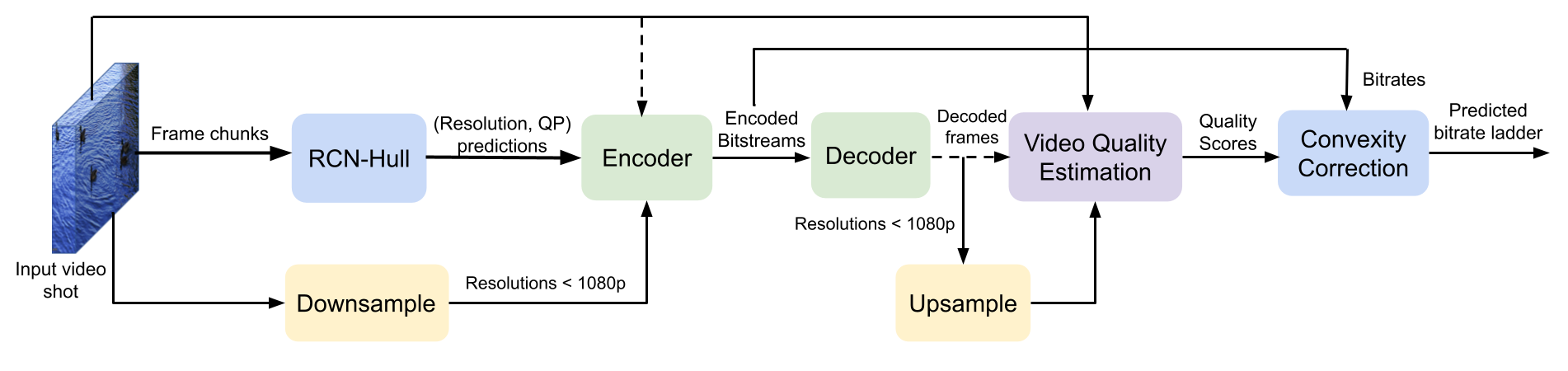}
    \begin{minipage}{0.5\textwidth}
    \caption{\small{{Schematic flow of our per-shot convex hull estimation method.}}}
   \end{minipage}
    \label{fig:overview}
\end{figure*}

Our present work is motivated by the remarkable success of deep learning based solutions to image/video processing tasks. As deep neural networks have been proven to be effective on numerous image and video analysis tasks, the merit of using data driven deep learning techniques to estimate the convex hull of possible encode configurations deserves exploration. However, the challenge of adopting this approach primarily arises from the sheer volume of video data that needs to be processed to perform the spatiotemporal scene complexity assessment on individual video shots. Unlike many video processing applications, where decisions can be made by either processing the frames sequentially or by processing short temporal chunks of a few frames at a time, the decision regarding the optimal points in the encode parameter space of a video shot requires a model to be able to assimilate relevant information over its entire duration. This presents formidable challenges with respect to memory and compute requirements, when processing high resolution video shots having arbitrary lengths. Convolutional neural network (CNN) based architectures that are commonly deployed for processing visual data have finite temporal receptive fields, which limits their ability to capture information over entire video shots of arbitrary lengths. While recurrent neural networks (RNNs) are conceptually suitable for processing sequential data due to their ability to retain past information in the form of long term memories, conventional RNN designs do not preserve the spatial structure of the input, which limits their success on visual data, where learning spatial structures is crucial. 

Towards addressing these problems, we have developed a deep learning based method that estimates the convex hull of video shots. Unlike existing machine learning (ML) based methods that aggregate handcrafted features to capture the spatial and temporal characteristics of video shots \cite{spatiotemporal, vmafladder}, we deploy a recurrent convolutional network (RCN) that jointly processes spatial and temporal information to provide learned spatiotemporal features. Our model is based on convolutional gated recurrent units (Conv-GRUs) \cite{cgru}, which make it possible to effectively capture the long term spatiotemporal properties over the durations of video shots, reminiscent of the incremental manner in which humans process information from videos. Further, unlike existing approaches \cite{spatiotemporal, vmafladder}, we formulate the prediction task as a multi-label classification problem rather than as a regression problem, which is learned by training on a large collection of lightly compressed, high quality videos from the publicly available Inter-4K dataset \cite{inter4K}, followed by fine tuning on a smaller collection of pristine, uncompressed videos, which are also publicly available. As compared to other alternative approaches developed for fast convex hull estimation, our method of joint space-time modeling and data driven learning delivers improved convex hull predictions on High Efficiency Video Coding (HEVC) \cite{hevc} encoded video shots, while also maintaining competitive advantages in terms of reductions in the time required to construct the convex hulls. Our convex hull estimation engine, which we refer to as RCN-Hull, is illustrated in Fig. 1. The chief contributions that we make are listed as follows:
\begin{enumerate}
\item we introduced a novel method of predicting the convex hulls used for per-shot video encoding using an RCN model that jointly processes spatial and temporal information
\item we constructed a large database of bitrate-quality (RQ) convex hulls on more than 500 video shots encoded by HEVC at 7 resolutions ranging from 1080p to 216p, and using 9 QP values.
\item we utilized a transfer learning based approach to mitigate challenges posed by the limited availability of raw video data on learning deep models for compression related tasks. 
\end{enumerate}

The rest of this paper is organized into five sections. In Section \ref{sec:literature} we review earlier work pertinent to the present topic. Section \ref{sec:database} describes the new database used to train the RCN model. Section \ref{sec:approach} elaborates our proposed convex hull estimation model (RCN-Hull). Experimental results are presented in Section \ref{sec:results}, and concluding remarks are given in Section \ref{sec:conclusion}.

\section{Related Work}
\label{sec:literature}
A comparative study of different bitrate adaptation strategies found that the perceptual impact of bitrate adaption in terms of Just Noticeable Difference (JND) scores is highly content dependent \cite{hlseval}. The impact of multiple factors such as bitrate adaptation algorithms, network conditions and video content in the context of adaptive video streaming was also explored in \cite{qoedb}, where human subjective quality of experience scores were collected to study these aspects. The per title encode optimization framework that was first introduced by Netflix \cite{pertitle} established the crucial role played by content complexity in determining bitrate ladders for adaptive streaming. The framework was further improved in \cite{pershot, do, pershot4K} by constructing the convex hulls of the RQ points at multiple resolutions and QP values on a per shot basis, which allowed them to achieve substantial BD-rate reductions. The process of constructing the convex hull of a video shot essentially consists of conducting an exhaustive search over the space of possible encoding parameters. This involves generating a pre-encode for each combination of the encoding parameters. The bitrates and video qualities obtained after the pre-encoding step determine the Pareto-efficient convex hull points in the encode parameter space. Most commonly, the encoding parameter space has two degrees of freedom, determined by the resolution and the encoding QP value. In a traditional, brute-force implementation of per-shot bitrate ladder selection, considerable computational resources are devoted to pre-encoding each video shot by varying both the resolution and the QP values.  It is also useful to vary other encoding parameters, such as the temporal framerate, when determining optimal bitrate ladders \cite{pstr}. However, adding more degrees of freedom inevitably causes a combinatorial increase in the complexity of the underlying search process. Moreover, even as streaming video resolutions continue increase over time, more complex video encoders become slower as they evolve to be able to achieve higher compression efficiency. Because of these two factors, the demands on computational resources for pre-encoding to construct content aware adaptive bitrate ladders continue to rapidly grow. Thus, development of systematic approaches to reduce this computational overhead has recently begun to attract research interest.

\begin{table*}[htb]
  \caption{Summary of the I-CV and UCV datasets.}
  \centering
  \begin{tabular}{|c|c|c|c|c|c|c|c|c|c|c|c|c|c|}
    \hline
    \multirow{2}{1.2cm}{\hfil \textbf{Dataset}} &  \multirow{2}{1cm}{\textbf{Purpose}} &  \multirow{2}{1cm}{\textbf{Quality}} &  \multirow{2}{1.2cm}{\textbf{Source resolution}} & \multirow{2}{1.4cm}{\textbf{Source framerate}} & \multirow{2}{1.4cm}{\textbf{Duration}} & \multicolumn{4}{c|}{\textbf{\# of video shots}}  \\
    \cline{7-10}
    & & & & & & Train &  Valid  & Test & Total  \\
    \hline
     \hfil I-CV & Training from scratch & High & UHD & 60 fps & 5s &468 & 58 &  54 & 580  \\ \hline
     \hfil UCV & Fine tuning &  Pristine & $\geq$ FHD  & 24 - 120 fps & 2.5s - 27.6s ($\mu$ = 9.86s, $\sigma$ = 4.79s) & 124 & 17 & 21 & 162 \\
\hline
  \end{tabular}
  \label{table:database}
\end{table*}

A just noticeable difference (JND) estimator was introduced for the purpose of conducting perceptual quality driven adaptive video streaming in \cite{jndestimator}. It used support vector regression (SVR) to predict JND scores with six input parameters (QP, bitrate, resolution, PSNR, SSIM and VMAF) obtained by performing curve fitting on a set of pre-encoded points. In \cite{opt-based}, the optimal design of encoding profiles for adaptive streaming was formulated as a non-linear constrained optimization problem, and solved using numerical programming techniques. A learning based, content aware adaptive bitrate ladder generation scheme was proposed in \cite{spatiotemporal}, which aggregated spatial features derived from the gray level co-occurrence matrix (GLCM) and temporal features that expressed temporal coherence (TC). The GLCM and TC features were used to analyze the spatial and temporal complexities of  video chunks, respectively, to predict the crossover QP points at which the RQ curves corresponding to different resolutions intersected. In \cite{proxy} it was shown that the convex hull of a video shot constructed with a faster encoder could be used as an effective proxy for the convex hull of the same shot constructed using a slower encoder having a higher compression efficiency. A reinforcement learning based method was used in \cite{deeprl}, which considered content characteristics, network capacity, and cloud storage costs, to estimate a proper encoder setting for each resolution. Yet, despite these promising efforts, the area of learning perceptually optimized convex hull to generate adaptive bitrate ladders has not been studied extensively in the open academic literature. 

Deep learning based techniques for video analysis have also advanced substantially in recent years. For example, CNNs, which can be used to analyze spatial data, and RNNs, which are quite useful for analyzing sequential data have been combined by stacking long short term memory (LSTM) layers onto CNN layers in \cite{rcn}. Promising results on diverse video analysis tasks such as activity recognition and video description were demonstrated using this approach. The idea was taken a step further in \cite{cgru}, where the convolutional layers were fused with gated recurrent units (GRUs) to form novel GRU-RCN layers, which are able to exploit the spatiotemporal structures of input video data, while also reducing the overall number of network parameters by spatially sharing parameters over the input. This architecture has been shown to achieve competitive performances on action recognition and video captioning tasks \cite{cgru}, as well as on video segmentation \cite{segment}.

\section{Convex Hull Database Construction}
\label{sec:database}
This section introduces the database of video shots that we collected from  publicly available source video sequences, along with the  convex hulls that we computed on them using HEVC encoding. Since there was no publicly available database of the sort, this step was indispensable to obtain the data needed to train a neural network model to learn efficient approximations of optimal convex hulls. 

\subsection{Source Content}
\label{subsec:source}
A significant challenge that researchers in the field of deep video compression grapple with is the paucity of publicly available pristine and uncompressed video content. Even the recently introduced BVI-DVC dataset for video compression \cite{bvidvc}, which compiled uncompressed video data from different publicly available sources, contains 280 unique contents. While this might be sufficient for training models that learn tasks on per frame or per chunk basis, it is inadequate for learning global tasks on entire video clips. We tackled this difficulty by adopting a two-step transfer learning based approach, whereby we first trained the RCN-Hull model on a larger, high quality, lightly compressed dataset derived from Inter4K  dataset \cite{inter4K}, then refined the model weights by fine tuning its later layers (those near the output) on a smaller set of pristine, uncompressed videos. The datasets used in each step are described in the following:
\begin{itemize}
\item \textbf{Inter4K-Compressed Video Set (I-CV)}: the Inter4K dataset \cite{inter4K} was introduced for video frame interpolation and super-resolution tasks. It consists of 1000 5-second UHD/4K video sequences captured at a frame rate of 60 frames per second (fps). The Inter4K database provides the video sources as high quality compressed videos. We visually inspected them and found them to be free of perceptual distortions. The clips in this dataset include wide ranges of challenging visual content, such as rapid motions, complex lighting, textures and many kinds of objects \cite{inter4K}. However, many of the video clips from this dataset contain shot changes. Since we are interested in predicting the convex hulls on a per shot basis, including shot changes is undesirable for our task. Hence, we manually inspected all 1000 clips to eliminate those containing any shot changes within their 5s duration. In the end, we obtained 580 video clips which we used to train the our RCN-Hull model from scratch. We will henceforth refer to this collection of data as the I-CV set. 
\item \textbf{Uncompressed Video Set (UCV)}: the goal of per shot encode optimization is to find the optimal encoding recipe for uncompressed source videos that will be streamed from the server. Although the videos in the I-CV set are of high perceptual quality, compression can still modify the underlying (bandpass) statistics of the videos. Since most successful video quality models such as VIF \cite{vif}, STRRED \cite{strred} and VMAF 
{} rely on natural video statistics models, it is possible that the prediction performance could be adversely affected when the RCN-Hull model is trained on the I-CV dataset, but deployed for estimating the convex hull of pristine, uncompressed videos. To compensate for this possibility, after pre-training on I-CV set, we used a smaller set of  uncompressed videos to fine tune the last few layers of the RCN-Hull model. We compiled a collection of 162 video shots from different publicly available contents \cite{uvg, sjtu, mcml, netflix-open, waterloo, bvi-sr, avt} that have been widely used for video compression research and benchmarking. We will refer to this assorted collection as the UCV set. The video sequences in the UCV set span multiple resolutions ranging  from FHD (1080p) to UHD/4K and includes videos of varying frame rates. Video sequences containing multiple shots were manually split into single shots. The duration of the video shots in the UCV set is also variable, and lies within the range of 2.5 - 27s, with mean ($\mu$) of 9.86s and standard deviation ($\sigma$) of 4.79s.  If multiple shots of a sequence were found to be visually redundant, only one of them was selected for inclusion in the database. Since the properties of the source contents in the UCV set are quite heterogeneous, we provide a detailed description of them in \cite{ucv}. 
\end{itemize}

The two datasets just described are summarized in Table \ref{table:database}, which also specifies the number of video shots included in the train, validation and test partitions of each set. All of the source videos of original resolutions higher than 1080p were downsampled to 1080p using Lanczos resampling with $\alpha = 3$. The I-CV dataset contains exactly 300 frames in each video shot. However, the UCV dataset contains video shots of varying lengths. Hence, we trimmed the shots that were assigned to  the training and validation partitions of the UCV dataset to 300 frames, making it possible to ensure consistent video batch sizes over all iterations during training. However, the videos in the test partition of the UCV dataset were not trimmed, and includes shots of variable lengths. 

\begin{figure}[t]
    \centering
    \subfloat[I-CV dataset]{%
       \includegraphics[width=0.98\linewidth]{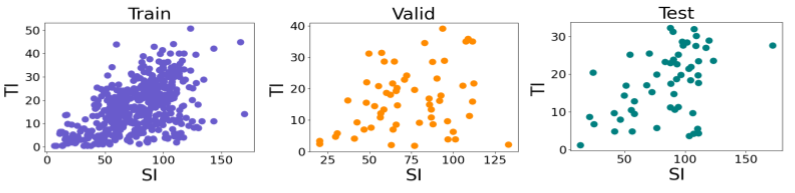}} \\     
  \subfloat[UCV dataset]{
   \includegraphics[width=0.98\linewidth]{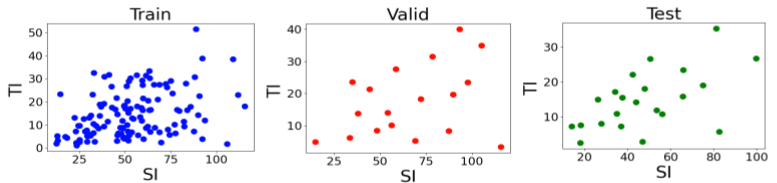}}
  \caption{\small{Scatter plots of SI and TI computed on the video shots from the database.}}
  \label{fig:SITI_plots} 
\end{figure}

We analyzed the content complexity of the video shots in the two datasets by computing the widely used spatial information (SI) and temporal information (TI) measurements on them \cite{siti}. Scatter plots of TI against SI are shown in Fig. \ref{fig:SITI_plots} for all subsets of I-CV and UCV. From Fig. \ref{fig:SITI_plots}, it can be observed that the SI values of the source videos in the I-CV set span a wider range of values than those of the source videos in the UCV set. By utilizing these two sources of data, our model was able to learn spatiotemporal scene complexity as well as capture the statistics of pristine videos, as later shown in Section \ref{subsec:prediction}.  

\subsection{Encoding}
We constructed ground truth bitrate ladders by encoding the source videos at 7 resolutions ranging from 1080p to 216p, which are typical of streaming applications \cite{encsettings}. The source videos were HEVC encoded at 9 QP values per resolution which ranged from 16 to 48, covering the range of visual qualities and bitrates attainable by varying the QP value over the allowable range of 0-51 for HEVC.  The specific set of resolutions was $\mathcal{P} = \{1920 \times 1080, 1280 \times 720, 960 \times 540, 768 \times 432,  640 \times 360, 480 \times 270, 384 \times 216 \}$, and the set of QP values we used was $\mathcal{Q} = \{16, 20, 24, 28, 32, 36, 40, 44, 48\}$. In principle, the sampling step size over the range of QP values should be determined by the JND levels of the encoded videos. However, since JND levels also depend on the content characteristics \cite{videoset}, a reliable model is needed to predict JND levels for different contents and spatial resolutions. Subjective JND datasets have been introduced for image and video coding standards such as JPEG \cite{jpegjnd} and H.264 \cite{h264jnd}. Based on these datasets, multiple JND models have been derived \cite{jndiq, gmm}. Since similar subjective JND measurements and corresponding prediction models for HEVC compressed videos are not publicly available, we used a constant step size of 4 to sample QP values in the range of 16-48. Using a smaller step size could potentially lead to a more efficient convex hull in terms of its rate-distortion performance, but would also yield more redundant points that are not separable in terms of JND levels. In fact, a subjective study on five different contents of diverse types found that the number of JND levels were between 2 and 6 when the contents were HEVC encoded (with x265) at a fixed resolution, with QPs sampled using a step size of 1 from the entire range of 0-51 \cite{jndtest}. Since our sampling step size produces more than 6 quality levels at each resolution over a smaller range of QPs, it can be expected to produce reasonably good coverage of the JND levels attainable within that range, while ensuring that the computational overhead of generating the pre-encodes for the large number of video contents present in the database remains tractable.

 Every source videos was encoded using the HM reference encoder at every possible combination $(p,q) \in \mathcal{P} \times \mathcal{Q}$. Thus, $|\mathcal{P}| \cdot |\mathcal{Q}|= 63 $ different bitstreams were encoded of each source content. Including all source video sequences drawn from both the I-CV and UCV sets, we generated a total of 46,746 different HEVC encoded videos.

\subsection{Convex Hull Generation and Representation}
\label{subsec:convex_hull}
Let $\bar{\mathbf{v}}$ be the frames of a video shot, and $\boldsymbol{f}_{pq}(\bar{\mathbf{v}})$ be a function that maps $\bar{\mathbf{v}}$ encoded at resolution $p$ and QP $q$ to the corresponding RQ point, denoted by $\boldsymbol{x}$. The set of RQ points is then given by $S = \Big \{ \boldsymbol{x} = \boldsymbol{f}_{pq}(\bar{\mathbf{v}}) \ | \ \boldsymbol{x} \in \mathbb{R}^{2}_{+} \Big \}$. Mathematically, the convex hull of the set $S$ is the set of all convex combinations of points in $S$, i.e.: 

\begin{equation}
\bar{S} = \Bigg\{ \sum_i \lambda_i\boldsymbol{x}_i \ | \ \boldsymbol{x}_i \in S, \lambda_i \geq 0 \ \text{and} \ \sum_i\lambda_i = 1 \Bigg \}.   
\end{equation}
However, in many practical contexts, such as video streaming, the term \textit{convex hull} is broadly used to refer to the boundary or envelope of the convex set $\bar{S}$ defined above. In accordance with this parlance, we also refer to the boundary of the smallest convex set that encloses the RQ points as the \textit{convex hull}. This boundary is given by:
\begin{equation}
\delta \bar{S} = \textbf{cl} \ \bar{S} \setminus \textbf{int} \ \bar{S} \text{,}
\end{equation}
where $\textbf{cl} \ \bar{S}$ and  $\textbf{int} \ \bar{S}$ refer to the closure and interior of the set $\bar{S}$, respectively, and are defined as:

\begin{equation}
\begin{split}
\textbf{int} \ \bar{S} &= \Big \{ \boldsymbol{x} \ | \  \exists \ \epsilon > 0 \ \text{s.t.} \ N_\epsilon(\boldsymbol{x}) \subseteq \bar{S}\ \Big \} \\
\textbf{cl} \ \bar{S} &= \mathbb{R}_{+}^{2} \setminus \textbf{int} \ (\mathbb{R}_{+}^{2} \setminus \bar{S}) \text{,}
\end{split}
\end{equation}
where $ N_\epsilon(\boldsymbol{x})$ is the $\epsilon$-ball centered at $\boldsymbol{x}$. A number of computational algorithms such as \cite{jarvis, graham, quickhull} have been developed to compute the points that constitute $\delta \bar{S}$. The points on $\delta \bar{S}$ are also a subset of the Pareto-efficient points of the RQ curve, which constitute the \textit{convex hull} of the corresponding video shot. 

As stated in Section \ref{sec:intro}, our convex hull prediction approach differs from current ML based methods like \cite{spatiotemporal, vmafladder}, since we pose the learning problem as multi-label classification instead of regression. Thus, we adopt a novel binary matrix representation of the convex hull, which can be learned using multi-label classification. Let $\mathcal{C}$ be the matrix of labels. Since the RQ curves were obtained by varying the resolutions $p$ and QPs $q$, we can index each RQ point on $\mathcal{C}$ using $p$ and $q$. Thus, we define:

\begin{equation}
\begin{split}
\mathcal{C}_{pq} & = 1 \ \text{if} \ \boldsymbol{f}_{pq}(\bar{\mathbf{v}}) \in \delta \bar{S} \\
& = 0 \ \text{otherwise}
\end{split}
\end{equation}

\begin{figure}[t]
\centering
    \includegraphics[width=8.8cm]{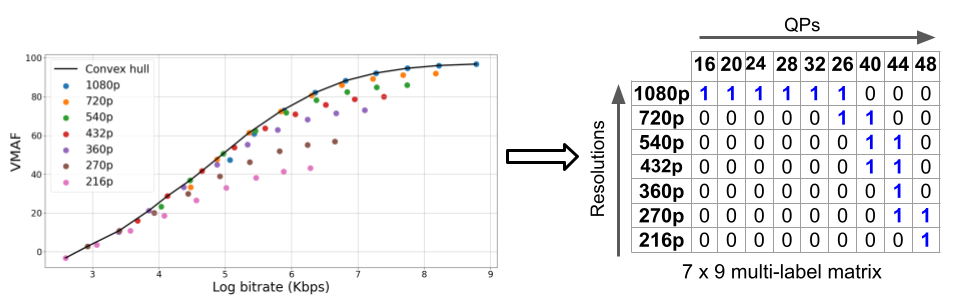}
    \caption{\small{{An example of a convex hull and its binary matrix representation for multi-label classification.}}}
    \label{fig:cvx_hull}
\end{figure}

In our dataset, $\mathcal{C}$ is a $7 \times 9$ matrix, corresponding to 7 resolutions and 9 QP values per resolution. This binary matrix is a concise representation of the convex hull $\delta \bar{S}$, and serves as the ground truth labels for the learning task. An example convex hull of a source sequence and the corresponding multi-label binary matrix is illustrated in Fig. \ref{fig:cvx_hull}\hspace{0.1cm}\footnote{\hspace{0.1cm}The envelope shown is non-convex because the bitrates are plotted in log scale to achieve separation of nearby points.}.

The encoded bitstreams were decoded and upsampled to the source resolution using Lanczos resampling with $\alpha = 3$. For each decoded and upsampled version,  the objective video quality with respect to the original source was predicted using VMAF \cite{vmaf, vmaf_neg}, which is a widely used full reference video quality model that has been extensively adopted by the streaming video and social media industries due to its ability to combine the features drawn from VIF \cite{vif} and DLM \cite{dlm} to deliver consistent quality prediction performance across different resolutions and genres. It is also computationally simpler as compared to other learned quality estimators that rely on deep neural networks. Next, the convex hull of the 63 RQ points  was computed for each source sequence using the quickhull algorithm \cite{quickhull}\hspace{0.1cm}\footnote{\hspace{0.1cm}We used the quickhull implementation available in the qhull library (http://www.qhull.org/).}. Using the same process, as described for VMAF, we also computed a second set of convex hulls using Multi-scale structural similarity (MS-SSIM) \cite{ms-ssim}, another perceptual model that is ubiquitously used for quality assessment.

Rather than providing information about individual points on the convex hull by formulating the ground truth as integer valued crossover QP points as in \cite{spatiotemporal, vmafladder}, our binary matrix formulation of the ground truth effectively conveys information about the entire RQ profile of a video shot to the model. Although the crossover QP points are critical for determining the bitrate ladder, information about the complete RQ profile can help the model construct better curve fits by learning useful relationships between neighboring points. Further, we have observed that the RQ curves corresponding to any two resolutions sometimes intersect at more than one point, and in such cases, the ground truth crossover QP cannot be unambiguously decided. The binary matrix representation also circumvents this issue, while providing a compendious representation of the convex hull.

\begin{figure}[htb]
    \centering
    \subfloat[Using VMAF \label{fig:vmaf_likelihoods}]{%
       \includegraphics[width=0.49\linewidth]{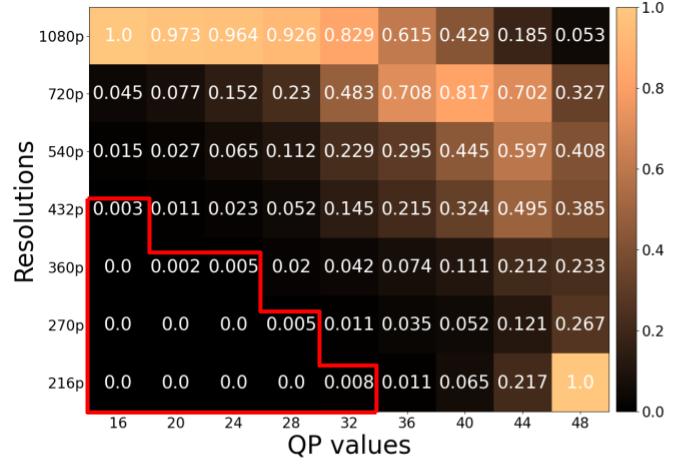}}     
  \subfloat[Using MS-SSIM\label{fig:msssim_likelihoods}]{
   \includegraphics[width=0.49\linewidth]{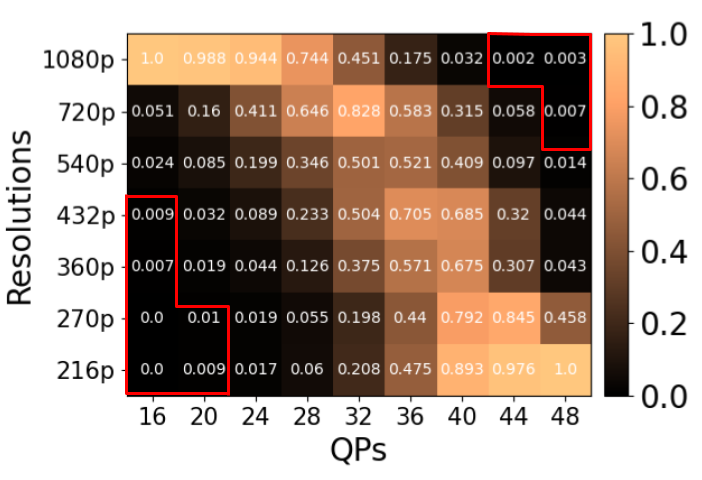}}
  \caption{\small{Statistical likelihood of each (resolution, QP) point's inclusion in the convex hulls.}}
  \label{fig:likelihoods} 
\end{figure}

The 63 RQ candidate $(p, q)$ points do not have the same likelihood of being included in the convex hull. Fig. \ref{fig:likelihoods} shows heatmap distributions of the statistical likelihood that $\mathcal{C}_{pq}=1$, computed on the dataset for each of the 63 candidate points using VMAF and MS-SSIM as quality metrics. From  Fig. \ref{fig:likelihoods} it is evident that the encodes corresponding to a few $(p, q)$ values were never included in the convex hulls, while a few others were rarely included. Thus, without loss of generality, we can eliminate the encodes at $(p,q)$ points where $\ell(\mathcal{C}_{pq}=1)$ is 0 or close to 0 from the set of candidate encodes, where $\ell$ represents the aforementioned statistical likelihood. Accordingly, we excluded those points having $\ell(\mathcal{C}_{pq}=1) \leq 0.01$ from the list of candidate encodes. The points thus eliminated are outlined by the red curves in Figs. \ref{fig:vmaf_likelihoods} and \ref{fig:msssim_likelihoods}, showing that the number of candidate points is reduced to 50 for the VMAF based convex hull and 54 for the MS-SSIM based convex hull, respectively. This reduction of the number of candidate encodes based on their statistical likelihoods of inclusion in the convex hull reduces the total time required by the baseline brute-force method of computing convex hulls, with negligible impact on the overall RD performance. 

To summarize, each sample of our dataset consists of video frames extracted from a single video shot, and the corresponding convex hull, representated as a $7 \times 9$ binary matrix. The ground truth binary representations of the convex hulls computed for each source sequence are provided in \cite{rcnhull}, along with instructions on how to download the corresponding source contents.

%

\begin{figure*}[htb]
    \centering
    \subfloat[Conv-GRU block\label{fig:model_a}]{%
       \includegraphics[width=0.32\linewidth]{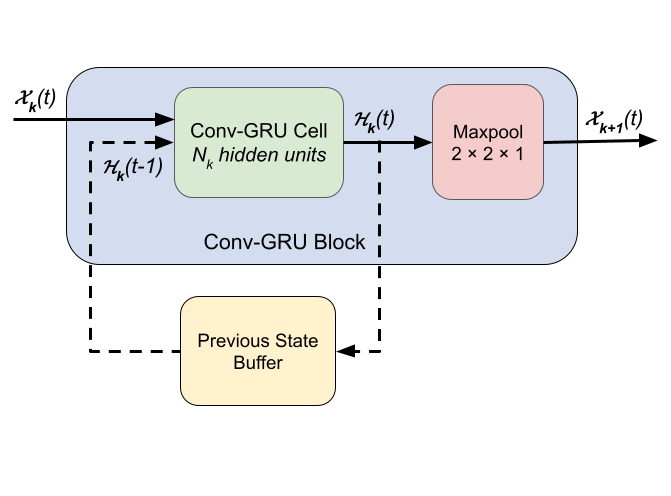}}     
  \subfloat[RCN-Hull model\label{fig:model_b}]{
   \includegraphics[width=0.68\linewidth]{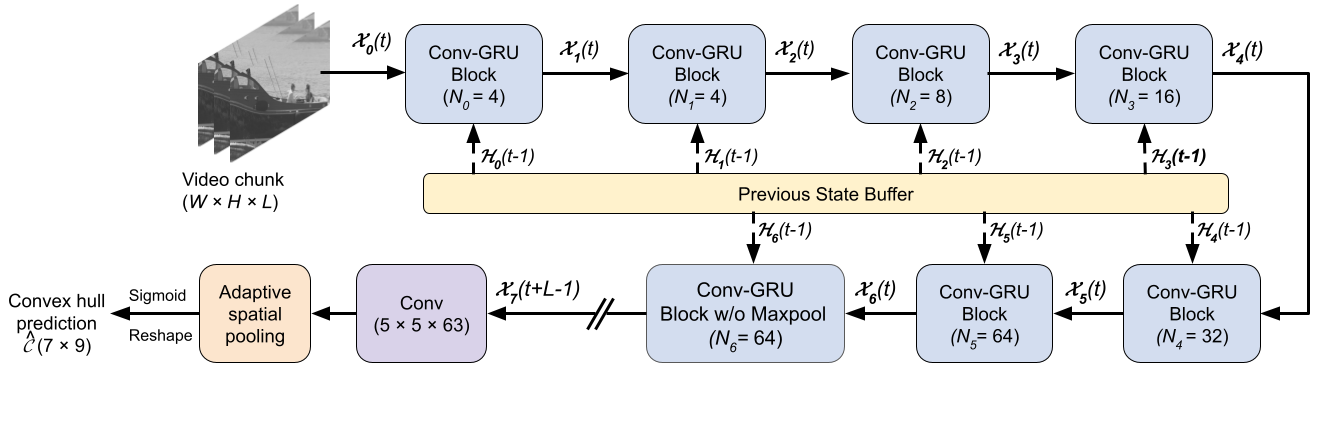}}
   
   \begin{minipage}{0.4\textwidth}
  \caption{\small{Architecture of the RCN-Hull model.}}
  \end{minipage}
  \label{fig:model} 
\end{figure*}

\section{RCN-Hull Model}
\label{sec:approach}

In this section, we describe the RCN-Hull model. First we provide the theoretical background of Conv-GRU layers, and discuss our motivation for using them. We then elucidate the model architecture that we developed for the purpose of convex hull prediction, as well as our training strategy for estimating convex hulls. 

\subsection{Conv-GRU: Background and Motivation}
The GRU is an efficient RNN architecture that achieves similar or better performance than LSTMs, while requiring fewer trainable parameters \cite{grueval}. However, since RNNs were originally envisaged as fully connected units, like other RNN architectures, fully connected GRUs also suffer from the drawback of not being able to capture the spatial structures of multidimensional inputs such as images or spatial feature maps. An intuitive solution that allows GRUs to learn spatial structures from input data is to replace the matrix multiplication operations used originally in case of fully connected GRUs with convolution operations, as suggested in \cite{cgru}, resulting in a recurrent unit, which we refer to as the Conv-GRU. 

Let  $\mathcal{X}(t)$ and $\mathcal{H}(t)$ represent the input and state information of the Conv-GRU unit at time step $t$. The Conv-GRU comprises three distinct gating mechanisms that performs the following operations \cite{cgru}:

\begin{subequations}
\small
\begin{align}
\mathcal{Z}(t) &= \sigma\big(W_z * \mathcal{X}(t) + U_z * \mathcal{H}(t-1) + B_z\big) \ \text{(update gate)} \\
\mathcal{R}(t) &= \sigma\big(W_r * \mathcal{X}(t) + U_r * \mathcal{H}(t-1) + B_r\big) \ \text{(reset gate)} \\
\hat{\mathcal{H}}(t) &= \sigma_h\big(W_h * \mathcal{X}(t) + U_h *\big(\mathcal{R}(t) \odot \mathcal{H}(t-1)\big) + B_h\big) \\
\mathcal{H}(t) &= \big(1 - \mathcal{Z}(t)\big) \odot \mathcal{H}(t-1) + \mathcal{Z}(t) \odot \hat{\mathcal{H}}(t)  \ \text{(output),}
\end{align}
\end{subequations}
where $W_z, W_r, W_h, U_z, U_r, U_h$ represent the weight parameters, and $B_z$, $B_r$, $B_h$ represent the bias parameters of the respective gates indicated by the subscript letters. $\sigma$ and $\sigma_h$ are the sigmoid and hyperbolic tangent activation functions, respectively. The convolution and Hadamard product operations are denoted by $*$ and $\odot$, respectively. $\mathcal{Z}(t)$ is the update gate vector which controls how much of the candidate activation of the current state, $\hat{\mathcal{H}}(t)$, contributes to the final current state information, relative to the state information received from the previous time step, $\mathcal{H}(t-1)$. The reset gate vector $\mathcal{R}(t)$ decides how much of the previous state's information is relevant, and can selectively ignore irrelevant information from previous time steps. 

Thus, in a Conv-GRU, the spatial structure of the input is preserved by the convolution operations, while the gating mechanism of the GRU components ensure that information from previous time steps are considered, and only relevant information is propagated to the next time step. This property makes Conv-GRUs a desirable choice for our use case, where the neural network needs to analyze spatiotemporal complexity over entire video shots when predicting their convex hulls. Since spatial structure of the input is preserved by convolution, a network made up of Conv-GRUs is capable of analyzing spatial complexity. It is also capable of incrementally analyzing spatiotemporal complexity, since it propagates the relevant information downstream as it processes frames sequentially. Since videos are characterized by a high degree of temporal redundancy, the network can make small updates to its state representation at every time step. 

Unlike contemporary architectures introduced for processing spatiotemporal data such as 3D CNNs \cite{3dcnn, C3D, I3D} and two stream CNNs \cite{twostream}, which have finite temporal receptive fields, Conv-GRUs are designed to effectively \textit{remember} information for any arbitrary number of future steps, as long as the information is still considered relevant by the training process, resulting in a temporal receptive field that effectively increases with time \cite{cgru}. While network architectures having finite  temporal receptive fields are suitable for processing short chunks of videos, it is not practicable to design models with receptive fields spanning hundreds of video frames, or even to process large numbers of frames (especially of high resolutions) as single inputs, due to obvious constraints on memory and computational resources. Thus, the elasticity of the temporal receptive field of the Conv-GRU is another factor that makes it a felicitous choice in our model design. It allows the network to sequentially consume short chunks of an input video, while retaining pertinent past information from all previously processed chunks, thereby keeping the memory usage within reasonable limits. The capability of Conv-GRU units to incrementally extract and selectively retain information thus mimics human processing of visual information from video data.

The decisive role of the Conv-GRU units in jointly modeling spatiotemporal content complexity is underscored by comparing its performance against an alternative model architecture, where the convolution and recurrent operations are decoupled, and performed sequentially, as discussed in Appendix A. 

\subsection{RCN-Hull Architecture}
The constituent units of the proposed RCN-Hull model are Conv-GRU blocks, where each Conv-GRU block is made up of a Conv-GRU cell and a spatial max pooling layer, as shown in Fig. \ref{fig:model_a}. At time step $t$, the input to the $k^{\text{th}}$ Conv-GRU layer is the output of the previous layer denoted by $\mathcal{X}_k(t)$, and the hidden state information of the corresponding layer at the previous time step $t-1$, which we denote by  $\mathcal{H}_k(t-1)$ and which is retrieved from the \textit{previous state buffer}. The convolution kernels used in all the Conv-GRU blocks of RCN-Hull are of size $3 \times 3$ and the $k^{\text{th}}$ layer has $N_k$ filters in each of its convolution kernels. The output of each Conv-GRU cell except the last one is spatially pooled  using a $2 \times 2$ max pooling window, and passed as the input to the next layer. At each time step, the previous state buffer is updated with the current hidden state representations of the Conv-GRU cells of all layers. 

The RCN-Hull model is composed of a series of seven Conv-GRU blocks, having 4, 4, 8, 16, 32, 64 and 64 hidden filter units in the convolutional kernel of each block from first to last, as shown in Fig. \ref{fig:model_b}. Since temporal neighborhoods of video frames usually contain highly redundant visual content, each video shot is processed with a temporal stride of $\delta$ frames. Thus, the luma channels of frames $0, \delta, 2\delta, 3\delta, \cdots$ are processed sequentially as short temporal chunks of length $L$ frames at a time. The inputs to the model for a video chunk starting at time step $t$ are the following:
\begin{itemize}
\item  $\mathcal{X}_0(t)$: a frame of size $W \times H$. 
\item $\mathcal{H}_0(t-1), \cdots, \mathcal{H}_6(t-1)$: the previous state features from the buffer. 
\end{itemize}
The hidden state information is initialized with zeros, i.e. at $t=1$, ${\mathcal{H}}_k(0) = \boldsymbol0 \text{,} \ \forall k$. The model produces its output after every $L$ frames are processed. After the final frame of the current chunk, given by $\mathcal{X}_0(t+L-1)$ is processed, the corresponding output of the final Conv-GRU layer is passed through a $5 \times 5$ convolutional layer with 63 filters, followed by an adaptive spatial pooling layer in order to have the desired output size. A sigmoid activation is applied on the final output to generate the prediction likelihoods, after which it is reshaped to a $7 \times 9$ matrix, yielding the convex hull prediction $\hat{\mathcal{C}}$.
Let $\Theta$ denote the set of all trainable parameters of the RCN-Hull model. The learning objective is then given by

\begin{equation}
\underset{\Theta}{\text{min}} \ \mathcal{L}(\mathcal{C}, \hat{\mathcal{C}}) \text{,}
\end{equation}
where $\mathcal{L}(\mathcal{C}, \hat{\mathcal{C})}$ is the cross-entropy loss. 

\subsection{Training Strategies}
We employed two intuitive strategies to efficiently train the RCN-Hull model, as described next. 

\textit{Incremental learning}: in order to maintain a tractable memory footprint, we adopted an incremental learning strategy, whereby, while training the RCN-Hull model, we processed small chunks of length $L$ frames from the input video shots and computed the binary cross entropy loss between the ground truth convex hull $\mathcal{C}$ and its prediction $\hat{\mathcal{C}}$. Although the loss was computed after processing every chunk, the corresponding gradients of the per-chunk losses were accumulated, and the model weights were updated only after the last frame of input video shot had been processed by the model. Due to the Conv-GRU based model design, newer information is acquired by the model as each chunk is processed, while relevant information from past chunks is retained in the hidden state representations learned, and propagated downstream. Intuitively, this strategy favors incremental learning, and allows the model to progressively improve its predictions with the availability of more information. At the same time, the gradient accumulation strategy imparts more stability to the training process. 
 
When using the RCN-Hull model for inferencing, each input video shot is simply passed to the model chunk by chunk, while the previous state buffer is updated with the learned intermediate hidden state representations. Since the model's output obtained after processing the last chunk of a shot effectively takes the entire video shot into account, we ignored the intermediate outputs and considered the output obtained after processing the last chunk of the video to be the final prediction. In Section \ref{subsec:prediction}, we demonstrate that this incremental learning strategy indeed allows the trained model to gradually improve its predictions while processing a video shot. 

\textit{Two step transfer learning}: as previously mentioned in Section \ref{subsec:source}, we used the I-CV and UCV sets to train the RCN-Hull model using transfer learning. First, we trained the entire model on the larger I-CV dataset. The purpose of this step was to provide the model with enough data to analyze and learn a wide range of content complexities. In the next step, we froze the weights of all but the last few layers, and updated the weights of the layers that were not frozen by training on the UCV set. This enabled the model to take into account the statistical structures of pristine videos. In Section \ref{subsec:prediction} we experimentally establish that this two-step transfer learning strategy improves the overall training performance. 

\subsection{Postprocessing}
\label{subsec:postproc}
During inference and deployment, the source videos were encoded at resolution $p \in \mathcal{P}$  and QP $q \in \mathcal{Q}$  if the corresponding RCN-Hull model prediction $\hat{\mathcal{C}}_{pq} = 1$. A convexity check was performed on the RQ curves obtained by encoding at the predicted set of $(p,q)$ points, and  the encodes corresponding to the points that do not lie on the convex hull of the RQ curves were eliminated, since these are likely to be sub-optimal. The points that pass the convexity check constitute the predicted approximations of the per-shot optimal bitrate ladders. 

\section{Experimental Results}
\label{sec:results}

We present the results obtained by evaluating the RCN-Hull model in terms of prediction performance and complexity reduction. We also compared its performance against three alternative approaches which are described next.  
\begin{itemize}
\item \textbf{Interpolation based hull (\textit{I-Hull})}: this simple approach reduces the computational overhead of generating a large number of pre-encodes when computing the optimal convex hull by encoding only a subset of the original set of QP points per resolution. The bitrates and quality values at those QP points that are not encoded are inferred using a suitable interpolation technique. We implemented the interpolation based approach by considering 5 of the 9 QP values in the set $\mathcal{Q}$ originally considered. The reduced set of QPs is thus $\mathcal{Q}^\prime = \{16, 24, 32, 40, 48\}$. At every resolution, the bitrates and quality values corresponding to the QPs in the set $\mathcal{Q} \setminus \mathcal{Q}^\prime $ were inferred using Piecewise Cubic Hermite Interpolation Polynomial (PCHIP) \cite{pchip}. The convex hulls of the RQ points comprising both actual and interpolated values of bitrates and qualities were then computed as the estimate of the optimal convex hull. If an interpolated point was found to lie on the hull, the corresponding encoding was performed to obtain the actual bitrate and quality value. 
\item \textbf{Proxy based hull (\textit{P-Hull})}: we implemented the proxy encoder based approach described in \cite{proxy} by encoding the source videos with the x265 encoder using the \textit{medium} preset at all 63 candidate $(p,q) \in \mathcal{P} \times \mathcal{Q}$ points. The convex hull generated using the bitrates and quality values of the encodes obtained with the faster x265 encoder was used as the proxy for the ground truth convex hull obtained for the reference HM encoder. 
\item \textbf{Feature based hull (\textit{F-Hull})}: we also implemented the handcrafted feature based approach proposed in \cite{spatiotemporal}, by extracting the suggested set of spatial features based on GLCM, temporal features based on TC, and spatial rescaling features given by the mean squared error between the original first frames of the video shots and their upsampled versions from the six lower resolutions. We used the Gaussian process classifer with rational quadratic kernel to learn the convex hulls using this set of features, as recommended in \cite{spatiotemporal}. The training partitions of the I-CV and UCV datasets were combined to create a larger collection of video shots, which was used to train the Gaussian process classifier. 
\end{itemize}
For each of the three alternative methods considered, a convexity check was performed on the predicted set of RQ points, and the points that failed the check were eliminated in the same way as described in Section \ref{subsec:postproc}. 


\subsection{Hyperparameters}


The RCN-Hull model depicted in Fig. \ref{fig:model} has 543,371 trainable parameters. While training, we used a temporal stride of $\delta = 5$ frames to process the video shot. The frames were consumed by the RCN-Hull model in chunks of size $L=3$ frames, which was limited by the available memory and batch size used. We used a batch size of 8 sequences. The RCN-Hull model was initially trained using the Adam optimizer \cite{adam} with a learning rate of $10^{-4}$ on the I-CV dataset. The number of layers selected for fine tuning was decided experimentally using cross validation, and the last two Conv-GRU blocks and the final $5 \times 5$ convolutional layer were fine tuned on the UCV dataset by reducing the learning rate to $10^{-5}$.

\subsection{Prediction Performance}
\label{subsec:prediction}
\begin{figure}[htb]
\centering
    \includegraphics[width=8.5cm]{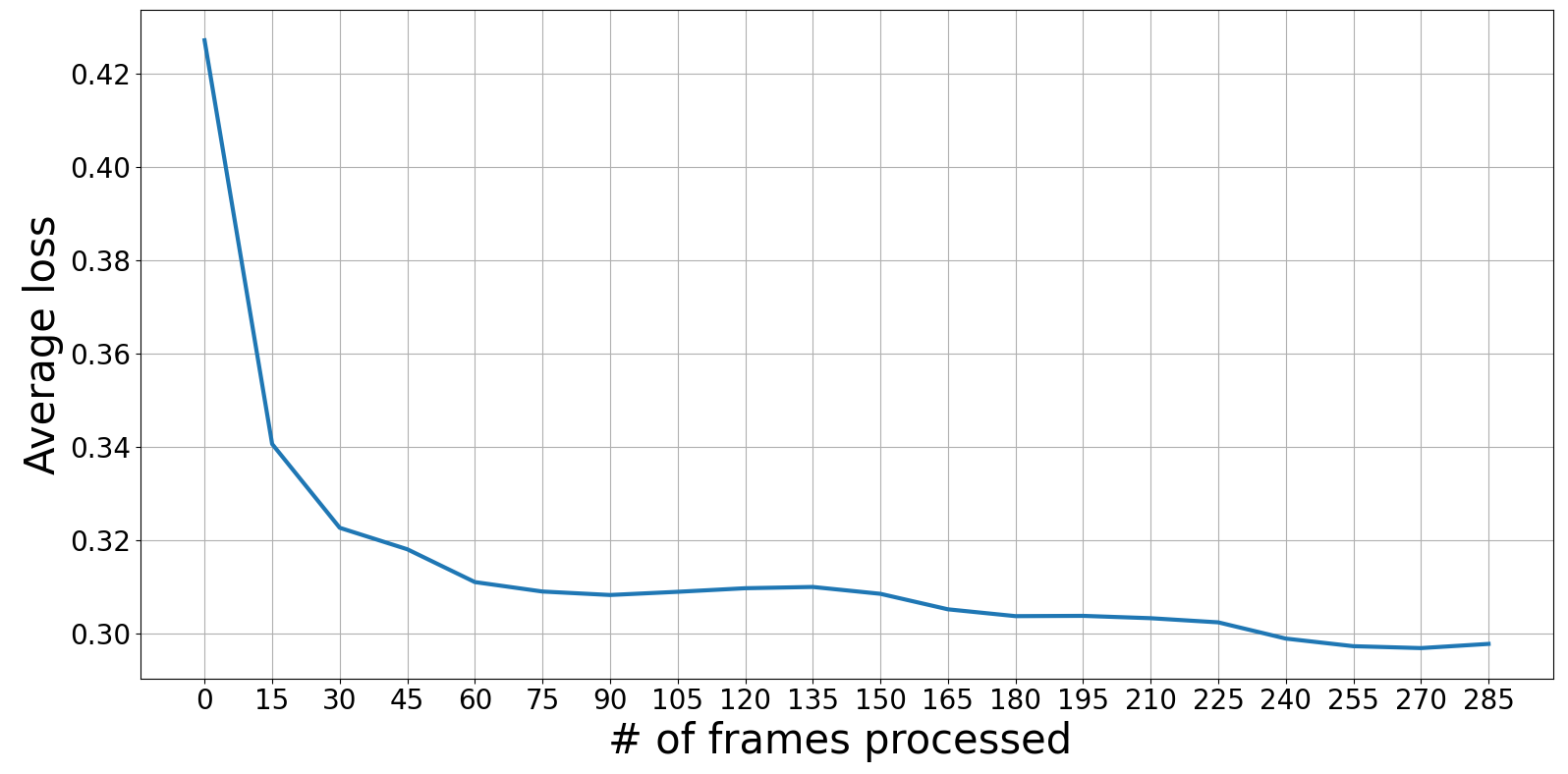}
    \caption{\small{{Plot of the average loss on the validation partition of the UCV dataset against the number of frames processed, using the trained RCN-Hull model.}}}
    \label{fig:loss}
\end{figure}

Since our target application is on uncompressed videos, we used the video shots from the UCV dataset for evaluation. We analyzed the the average validation loss while processing the chunks of the video shots sequentially with the RCN-Hull model trained on the I-CV and UCV datasets using the two-step transfer learning approach. The trained model's loss  on the validation partition of the UCV set as a function of the number of frames processed is plotted in Fig. \ref{fig:loss}. When interpreting Fig. \ref{fig:loss}, it should be noted that  the loss was obtained with the trained model with fixed parameters. From Fig. \ref{fig:loss} shows that the average loss decreases progressively as more frames are processed by the model. Since the weights of the model were fixed, it can be inferred that the trained model was able to gradually improve its predictions by reducing the associated uncertainties, as it ``\textit{watched}" the videos shots, corroborating our earlier claim that the Conv-GRU based architecture is efficacious at extracting and retaining critical visual information to develop a holistic understanding of the spatiotemporal complexity of each video shot. 

 \begin{table}[htb]
\caption{Classification performance metrics on the UCV test set and their 95\% CIs.}
\centering
\vspace{0.1cm}
 \begin{tabular}{|c | c | c | c |} 
 \hline
  \multirow{2}{1cm}{\textbf{Approach}} &  \multirow{1}{1cm}{\textbf{Precision}}
  & \multirow{1}{1cm}{\hfil \textbf{Recall}} & \multirow{1}{1.1cm}{\hfil \textbf{F1 score}} \\
& \textbf{(\%)} & \textbf{(\%)} & \textbf{(\%)} \\
\hline
 \multirow{2}{1.7cm}{F-Hull \cite{spatiotemporal}} & 88.6  & 66.1 &  75.8 \\ 
 & [87.9, 89.8] & [65.6, 68.0] & [75.2. 77.4] \\
 \hline
\multirow{2}{2.7cm}{RCN-Hull (I-CV only)}  & 89.3  & 75.1 &  81.6 \\ 
 & [88.3, 90.0] & [73.6, 75.9] & [80.3, 82.3] \\
\hline
\multirow{2}{2.7cm}{RCN-Hull (UCV only)}   & 88.9  & 70.1 &  78.4 \\ 
 & [87.4, 89.1] & [69.2, 71.1] & [77.3, 79.0] \\
\hline
\multirow{2}{2.8cm}{RCN-Hull (I-CV+UCV)}   & \textbf{91.6}  & \textbf{77.8} & \textbf{84.1}   \\ 
& [90.6. 91.9] & [76.0. 77.9] & [82.7, 84.3]
 \\
\hline
\end{tabular}
\label{table:metrics}
\end{table}

\begin{table*}[htb]
\caption{BD-rates and complexity Reduction of RCN-Hull on the UCV test set}
\centering
\vspace{0.1cm}
 \scalebox{0.86}{
 \begin{tabular}{|c | c | c  c |c c|c c|} 
 \hline
  \multirow{2}{1cm}{\textbf{Sequence}} &  \multirow{1}{0.8cm}{\textbf{\# of frames}}
  & \multicolumn{2}{c|}{\textbf{BD-rates (\%)}} & \multicolumn{2}{c|}{\textbf{Time savings(\%)}} & \multicolumn{2}{c|}{\textbf{Reduction in \# of encodes (\%)}} \\
&  & \textbf{ VMAF} &   \textbf{MS-SSIM} & \textbf{VMAF} & \textbf{MS-SSIM} & \textbf{VMAF} & \textbf{MS-SSIM}  \\
\hline
 \textit{Touchdown pass}  & 570  & -0.03 & 0.12 & 58.9  & 57.4  & 60.5 & 59.2 \\ 
  \textit{Rainroses}  & 500  & -0.13 &  0.91  & 57.5 & 50.9 & 62.9 & 64.8 \\ 
   \textit{Pedestrian area}  & 375  & 0.0 & 0.01 & 36.9 & 49.3 & 56.8 & 61.1 \\  
   \textit{Into tree}  & 500  & -0.01 & -0.18 & 57.2 & 50.1  & 55.9  & 70.8 \\ 
   \textit{Blue sky}  & 217  & 2.67 & 0.55 & 48.4 & 50.7 & 70.4 & 66.6 \\ 
   \textit{Runners}  & 300  & 0.24 & -0.16 & 50.8 & 53.3 & 62.2 & 68.7\\ 
    \textit{Race night}  & 600  & -0.13 & -0.07 & 59.7 & 58.7 & 61.4 & 59.2 \\ 
    \textit{Netflix Tango}  & 294  & 0.0 & 0.05 & 56.1 & 49.8  & 65.8 & 59.2\\ 
    \textit{Netflix Ritual Dance}  & 424  & 0.01 & 0.02 & 58.6 & 60.7 & 60.5 & 68.5 \\ 
    \textit{Netflix Crosswalk}  & 300  & -0.04 & 0.13 & 48.7 & 55.1 & 55.3 & 57.4 \\ 
    \textit{Netflix Bar Scene}  & 360  & 0.16 & 0.15 & 51.2 & 51.0  & 58.1 & 55.5 \\ 
  \textit{Netflix Driving POV}  & 300  & -0.19 & 0.01 & 58.4 & 57.3 & 66.7 & 63.0\\ 
  \textit{Light rail}  & 300  & -0.12 & 0.77 & 50.7 & 43.8  & 76.9 & 66.0 \\ 
  \textit{Halftime music}  & 300  & 0.27 & -0.02 & 55.5  & 47.6  & 70.3 & 66.7\\ 
   \textit{Flags}  & 330  & -0.86 & 0.20 &  53.4 & 57.0 & 63.4 & 61.1 \\ 
    \textit{Fountains}  & 300  & 0.57 & 2.15 & 61.0 &  61.5 & 68.7 &  70.4 \\ 
    \textit{GTAV scene}  & 691  & 2.03 & 1.50 & 47.0 & 38.4 & 64.9 & 73.5\\ 
    \textit{Flower Focus}  & 300  & -0.41 & -0.03 & 44.7  & 58.3  & 46.5 & 61.1\\ 
   \textit{Beauty}  & 600  & 1.45 & 0.75 & 64.5 & 53.9 & 50.0 & 57.4 \\   
    \textit{Flowers}  & 300  & -0.36 & 0.08 & 56.1 & 63.7 & 61.8 & 68.5 \\ 
    \hline
    \multicolumn{2}{|c|}{Average} & 0.26 & 0.35 & 53.8 & 53.4 & 62.0 & 63.9 \\ 

\hline
\end{tabular}
}
\label{table:performance}
\end{table*}
We evaluated the performance of the trained and fine-tuned RCN-Hull model on the test partition of the UCV dataset. Since the ground truth convex hull matrices contain many zeroes, classification accuracy is not a suitable performance indicator. Hence, we report the prediction performance using the precision, recall and F1 score, which are standard classification metrics that provide a more meaningful assessment of a model's performance in the presence of an imbalanced class distribution. The average precision, recall and F1 score using the proposed two-step transfer learning approach were 91.6\%, 77.8\% and 84.1\%, respectively as given in Table \ref{table:metrics}. We also report the corresponding metrics obtained without transfer learning in Table \ref{table:metrics}. It may be seen that both the precision and recall metrics, and consequently the F1 scores, were lower when the model was individually trained using either the I-CV or the UCV dataset, which validates the efficacy of the transfer learning strategy in improving the model's performance. Also, the metrics for the feature based ML method \cite{spatiotemporal} given in the first row of Table \ref{table:metrics}, show that it is outperformed by RCN-Hull in terms of all three metrics, not only using the transfer learning approach, but also when trained exclusively on either dataset. Table \ref{table:metrics} also reports the 95\% confidence intervals (CIs) of each metric, computed using 1000 bootstrapped samples of the data.  

Table \ref{table:performance} lists BD-rates of the RQ curves generated using the RCN-Hull model predictions with the optimal ground truth convex hulls of the test sequences used as the reference. The PCHIP \cite{pchip} interpolation method which has been widely employed to compute BD-rates in codec standardization efforts due to its relative stability \cite{bjontegaardbible}, was used as the interpolation method to compute the BD-rates reported here. Even though the lengths of the video shots used during training were fixed at 300 frames, the test sequences were of different lengths, as specified by the second column of Table \ref{table:performance}. The attained performance on shorter video chunks having fewer than 100 frames is presented in Appendix \ref{app: appb}. Since the convex hull is the optimal RQ curve, the negative BD-rates in Table \ref{table:performance} might seem counterintuitive. However, in those cases where the number of predicted points in $\hat{\mathcal{C}}$ are fewer than the number of points on the ground truth convex hull $\mathcal{C}$, the computed BD-rate can be negative, due to the logarithmic transformation applied to the bitrates for BD-rate calculation, as demonstrated in Appendix \ref{app: appc}. Thus, it is desirable to have the BD-rate magnitudes close to 0, signifying small deviations from the optimal convex hull. Since the prediction errors produced both positive and negative BD-rates, in this instance the average of the signed BD-rates is not a reliable performance indicator. For this reason, we instead use the more indicative dispersion of BD-rates around 0. Table \ref{table:performance} shows that the BD-rate magnitudes obtained via RCN-Hull are small for most of the sequences, with only 3 test sequences having a BD-rate magnitude greater than 1\%.  The average BD-rate obtained by using RCN-Hull to predict the convex hulls  was 0.26\%, and the mean absolute deviation (MAD) was 0.57\%, when VMAF was used to estimate the video qualities for convex hull computation. The VMAF values were restricted in the range of [21, 99] when computating the BD-rates reported here. VMAF values below 21 correspond to encodes of low visual quality which are typically not suitable for streaming applications, and thus we excluded them from the performance evaluation. Also, since VMAF values saturate above 99, streaming multiple encodes having VMAF values greater than 99 adds little practical benefit from a perceptual quality standpoint, which justifies our choice of the upper boundary of the VMAF range. The same range of VMAF values was also used to compute the complexity reduction values reported in the next section. 

\begin{figure*}[htb]
    \centering
    \subfloat[BD-rate distribution \label{fig:bdrate}]{%
       \includegraphics[width=0.49\linewidth]{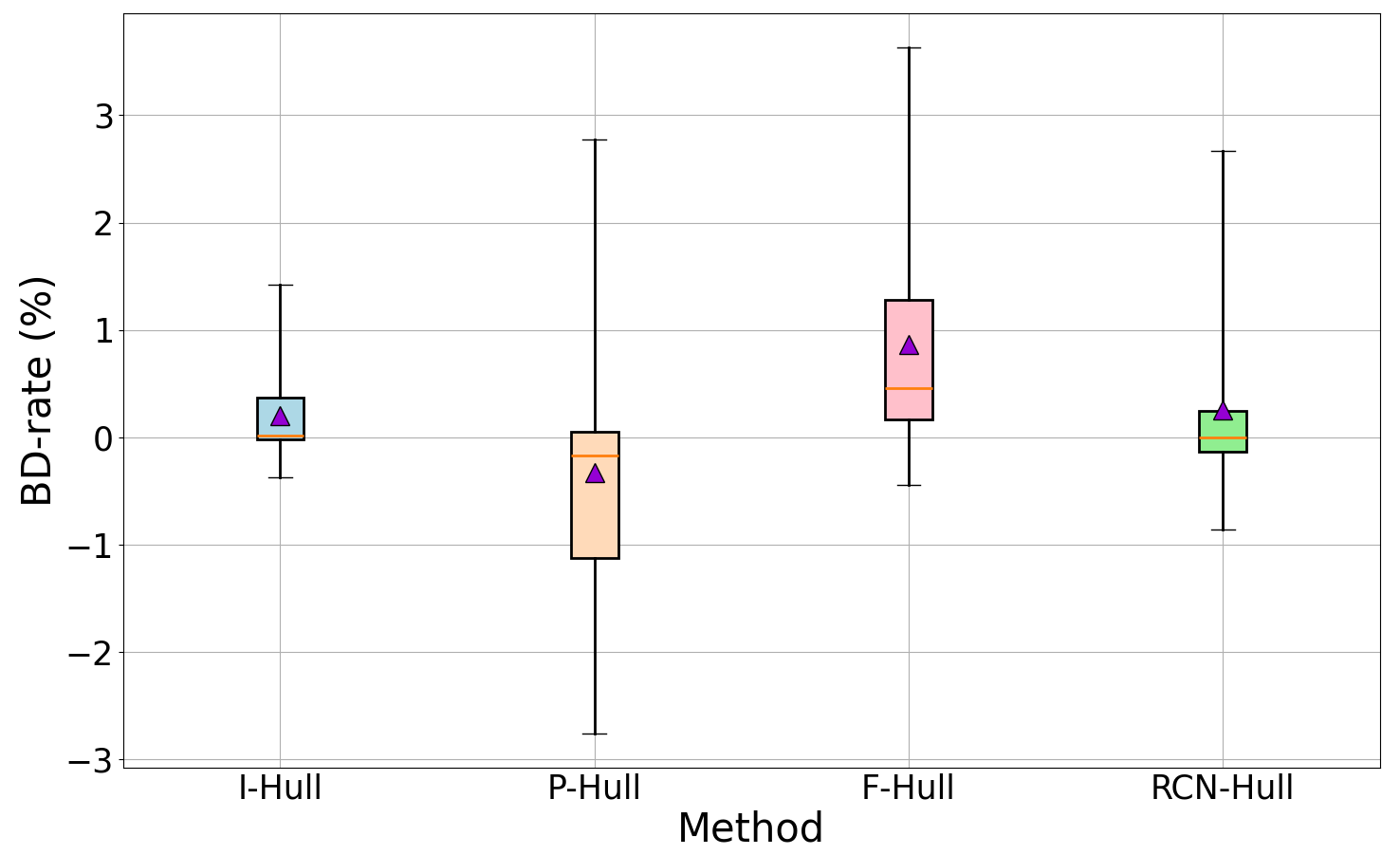}}     
  \subfloat[Time savings distribution\label{fig:speed}]{
   \includegraphics[width=0.49\linewidth]{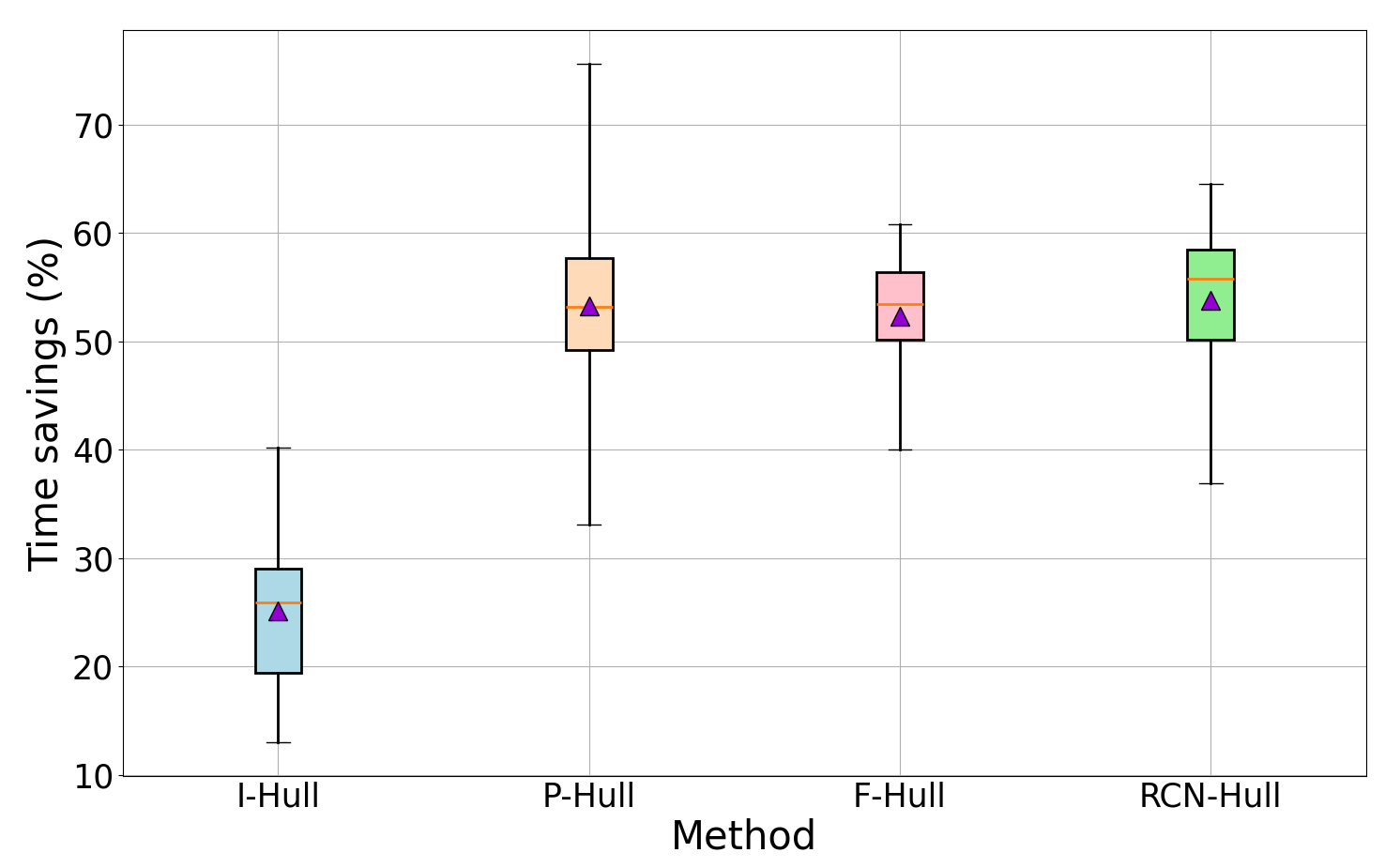}}
  \caption{\small{Box plots of (a) BD-rate and (b) time savings for each of the compared methods: I-Hull, P-Hull \cite{proxy}, F-Hull \cite{spatiotemporal} and RCN-Hull.}}
  \label{fig:comparison} 
\end{figure*}

\subsection{Complexity Reduction}
\label{subsec:complexity}
A total of 50 candidate encodes as given by Fig. \ref{fig:vmaf_likelihoods} need to be generated to produce the convex hulls obtained with VMAF as the quality metric over the 7 resolutions and 9 QP values. However, RCN-Hull only requires the model inference on the video shot and generation of those encodes that are predicted to lie on the convex hull. The number of encodes required while using RCN-Hull is thus $\underset{p \in \mathcal{P}}{\sum}\underset{q \in \mathcal{Q}}{\sum} \hat{\mathcal{C}}_{pq}$, which implies significant reduction of the complexity of bitrate ladder construction. Columns 5-8 of Table \ref{table:performance} report the reduction in the time required to construct the convex hulls of the test sequences obtained using RCN-Hull, and the reduction in the number of encodes required for this purpose, respectively. These values were computed taking into account even the points removed in the postprocessing step of Section \ref{subsec:postproc}, since every predicted point in $\hat{\mathcal{C}}$ were encoded before the convexity correction step, regardless of their inclusion in the final bitrate ladder. The average reduction in the number of encodes was 62.0\%, while the average time savings was 53.8\%, as reported in the bottom row of the table for the prediction of convex hulls derived using VMAF. Of the 50 candidate encodes for the VMAF-based convex hulls, only 18.15 belonged to the ground truth convex hulls on average, while the average number of encodes required per video shot was reduced from 50 to 18.60 using RCN-Hull.

The CPU  inference time of our model was 5.91s per 3-frame chunk. Assuming a temporal stride of $\delta = 5$ frames, the total CPU inference time of a 300 frame video shot is about 120s, which is negligible as compared to the overall encoding time, which is of the order of $10^4$s for the same number of frames. 

\begin{figure}[htb]
\centering
    \includegraphics[width=8.5cm]{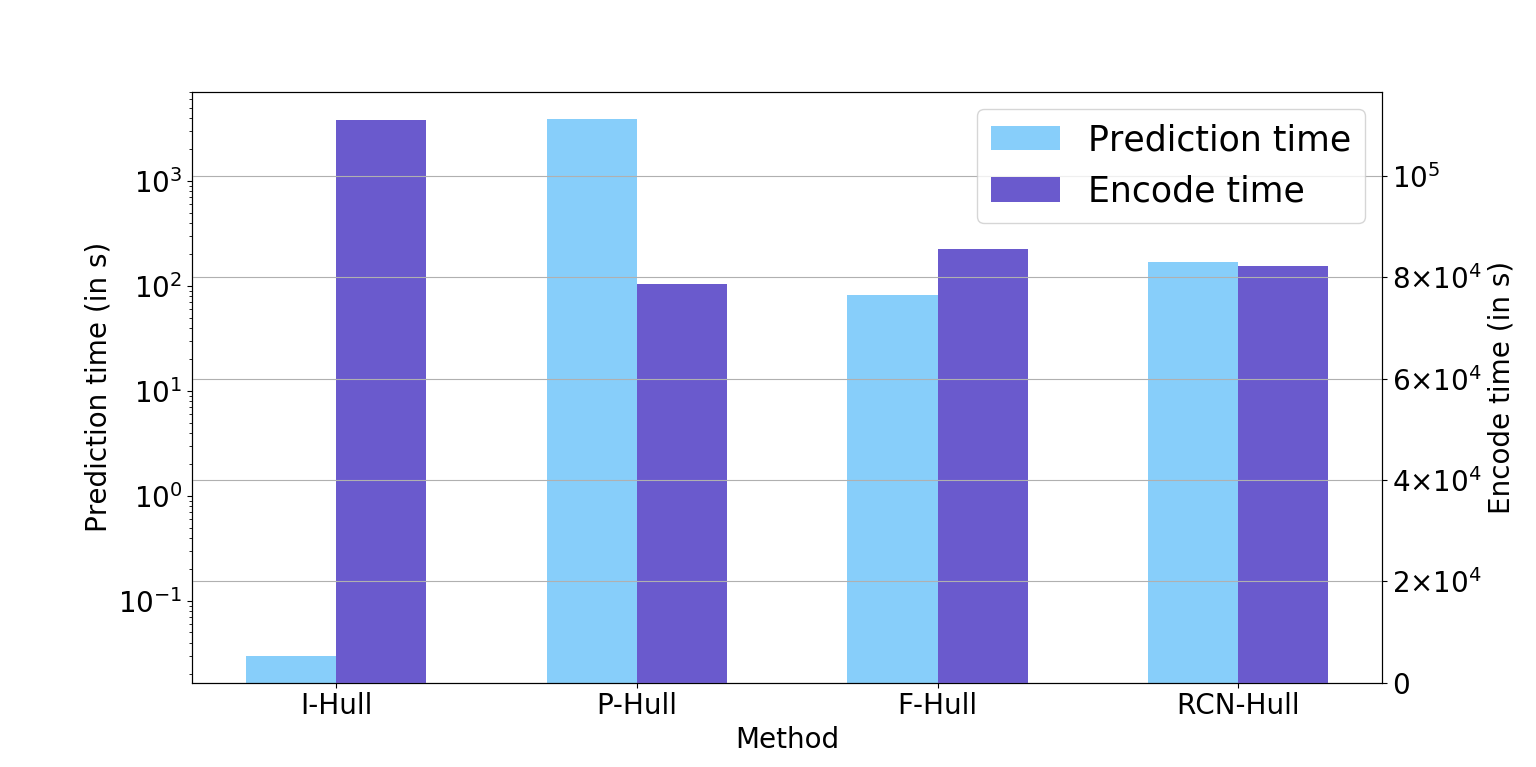}
    \caption{\small{{Prediction times and encode times of I-Hull, P-Hull \cite{proxy}, F-Hull \cite{spatiotemporal} and RCN-Hull.}}}
    \label{fig:time_barplot}
\end{figure}


The RCN-Hull model is also applicable when other quality metrics are chosen to construct the convex hulls. To demonstrate this point, we supplemented Table \ref{table:performance} with the results obtained by training the model on the convex hulls obtained using the MS-SSIM quality model. In this case, there are 54 candidate encode points, as shown in Fig. \ref{fig:msssim_likelihoods}. The MS-SSIM scores were converted to the dB scale ($\text{MS-SSIM}_{\text{dB}}= -10\log_{10}(1-\text{MS-SSIM})$), and bounded within the range [7-25] dB which corresponds to a similar practical quality range, as explained in Section \ref{subsec:prediction}. Table \ref{table:performance} shows that our proposed model attained similar levels of RD performance  (columns 3-4) as well as computational performance (columns 5-6 and columns 7-8) when using both VMAF and MS-SSIM as the underlying perceptual quality model.

\begin{table*}[htb]
\caption{Performance summary of the compared convex hull estimation methods.}
\centering
\vspace{0.1cm}
\scalebox{1.0}{
 \begin{tabular}{|c | c | c | c | c |} 
 \hline
 \textbf{Performance measure} & \textbf{I-Hull} & \textbf{P-Hull} \cite{proxy} & \textbf{F-Hull} \cite{spatiotemporal} & \textbf{RCN-Hull} \\
\hline
  \multirow{2}{3cm}{Average time savings (\%)} & 25.1  & 53.2 &  52.4  & 53.8\\
  & [21.9, 28.2] & [49.5, 56.9] & [49.8, 54.9] & [50.8, 56.5] \\
 \hline
  \multirow{2}{4cm}{Average BD-rate magnitude (\%)} & 0.27  & 1.03 & 0.91  &  0.48\\
 & [0.11, 0.44] & [0.61, 1.49] &  [0.44, 1.51] & [0.19, 0.82] \\
 \hline
 \multirow{2}{2.5cm}{Average BD-rate (\%)} & 0.20  & -0.32 & 0.83  &  0.26\\
 & [0.01, 0.39] & [-0.92, 0.26] &  [0.29, 1.48] & [-0.13, 0.64] \\

 \hline 
 MAD of BD-rate (\%) & 0.31  & 0.99 &  0.96  & 0.57\\ 
 \hline
 SD of BD-rate (\%) & 0.43  & 1.43 &  1.37  & 0.84\\ 
 \hline

\end{tabular}
}
\label{table:summary}
\end{table*}

\begin{figure*}[htb]
\centering
\subfloat
    {\includegraphics[width=0.32\textwidth]{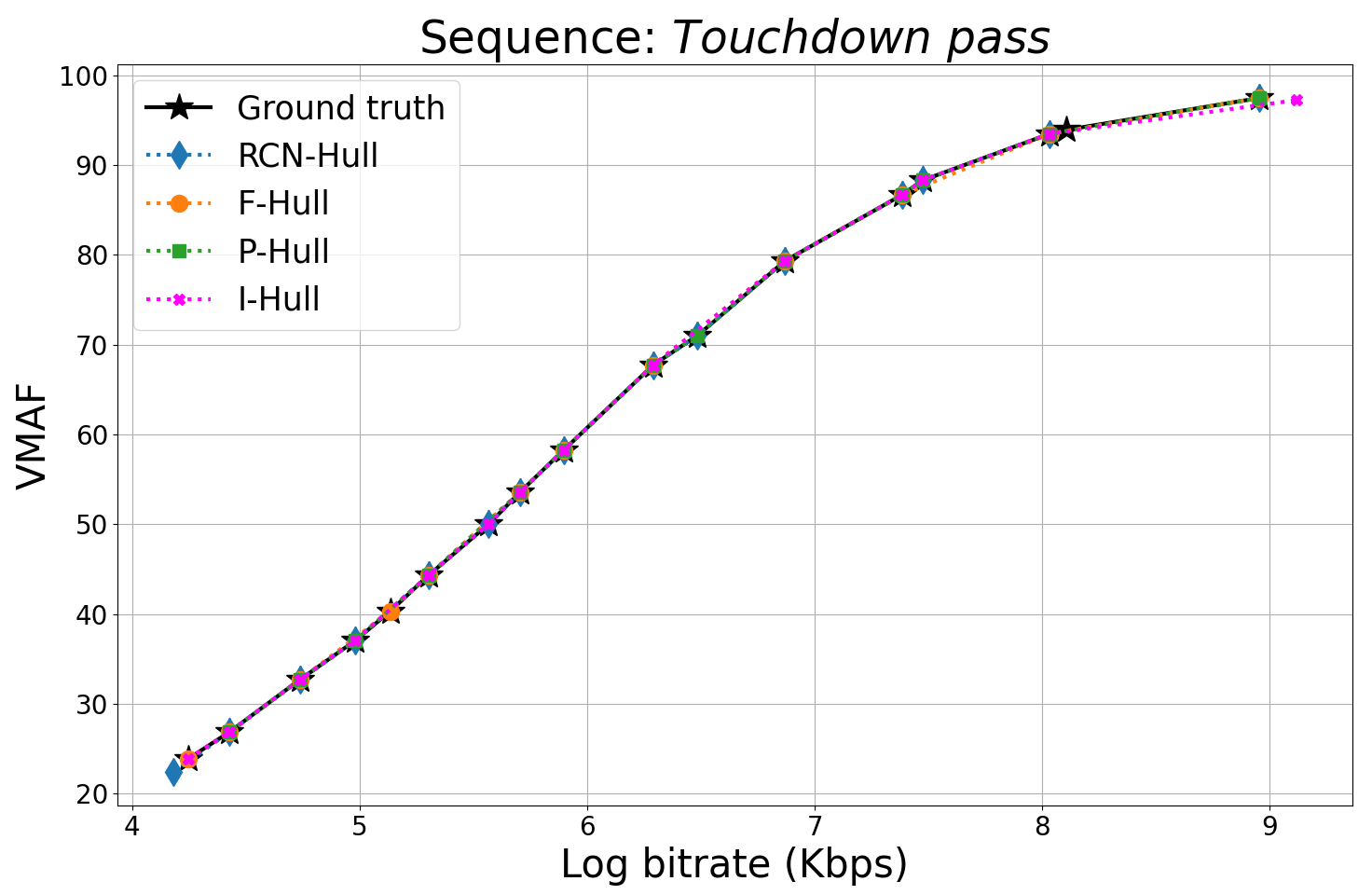}}
\hfil
\subfloat
    {\includegraphics[width=0.32\textwidth]{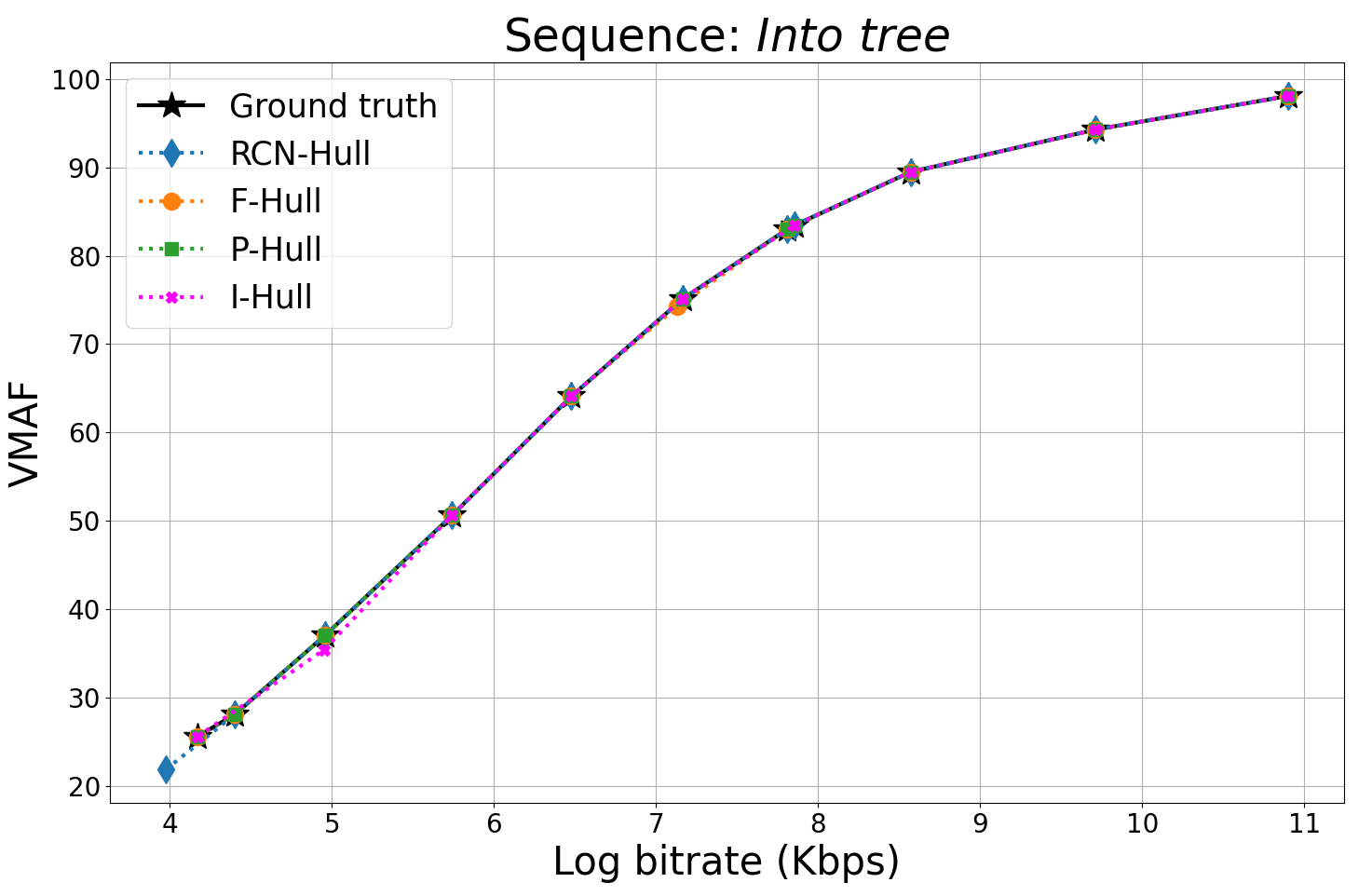}}
\hfil
\subfloat
    {\includegraphics[width=0.32\textwidth]{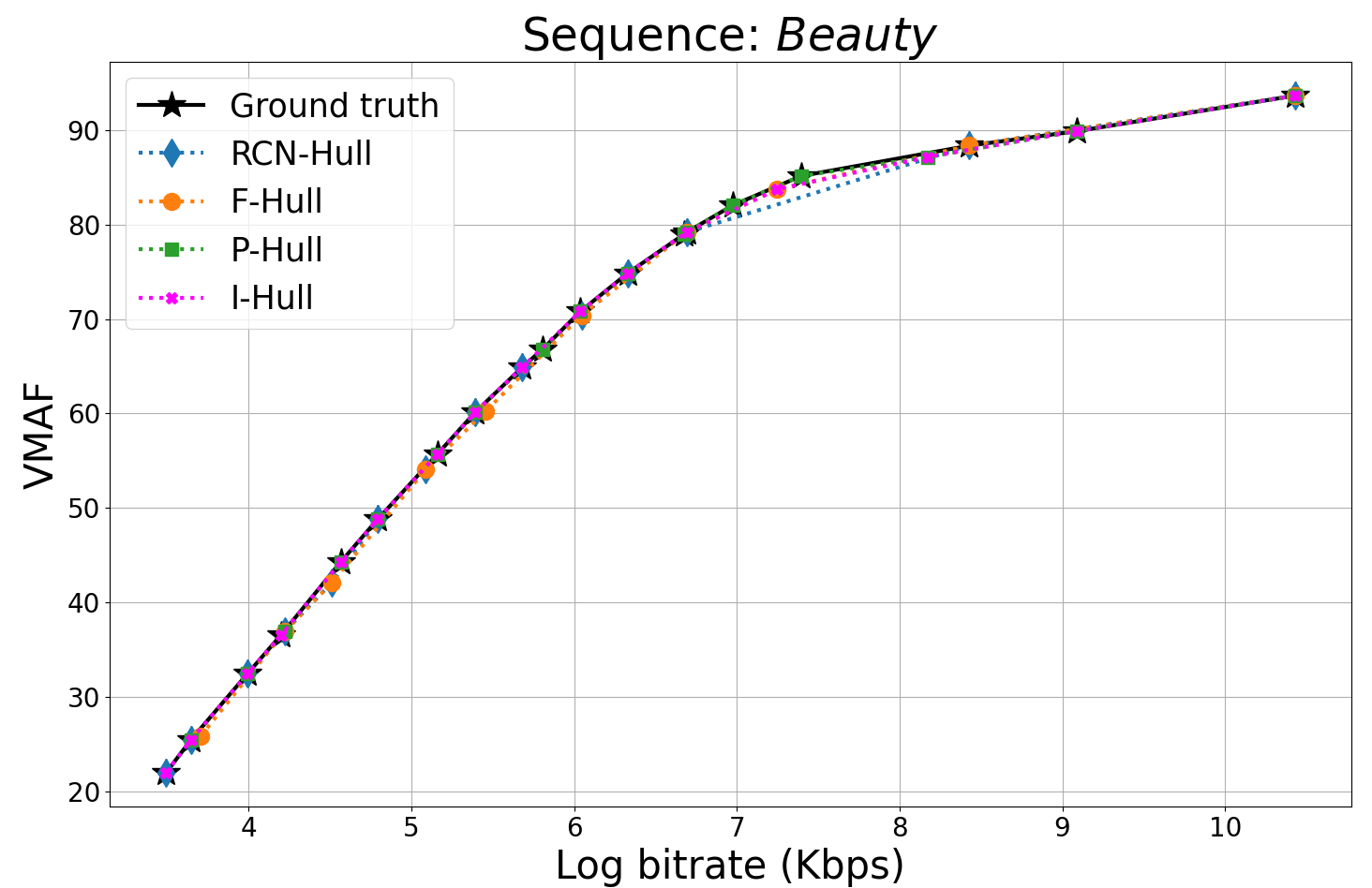}}
\hfil
\subfloat
    {\includegraphics[width=0.32\textwidth]{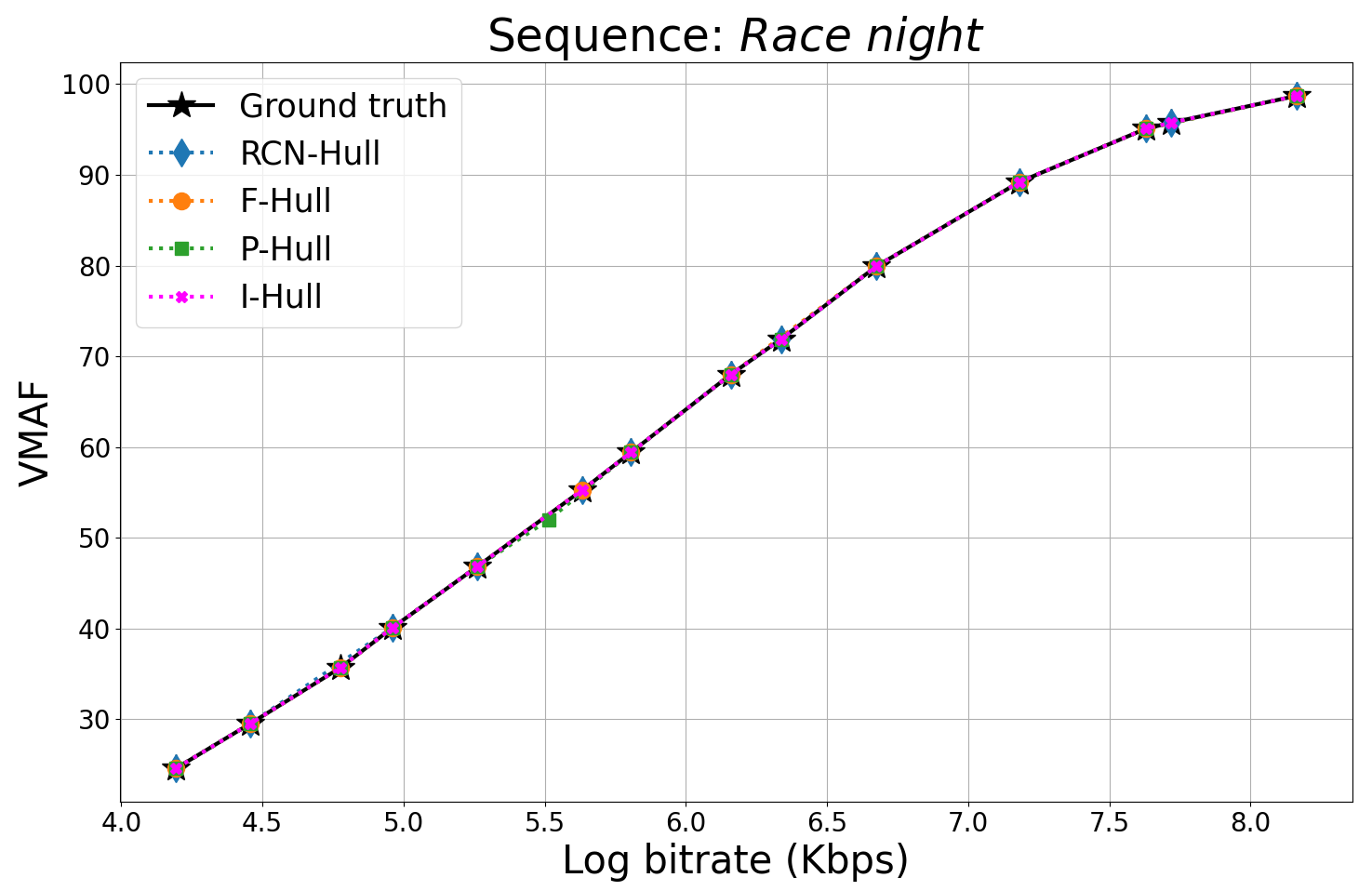}}
\hfil
\subfloat
    {\includegraphics[width=0.32\textwidth]{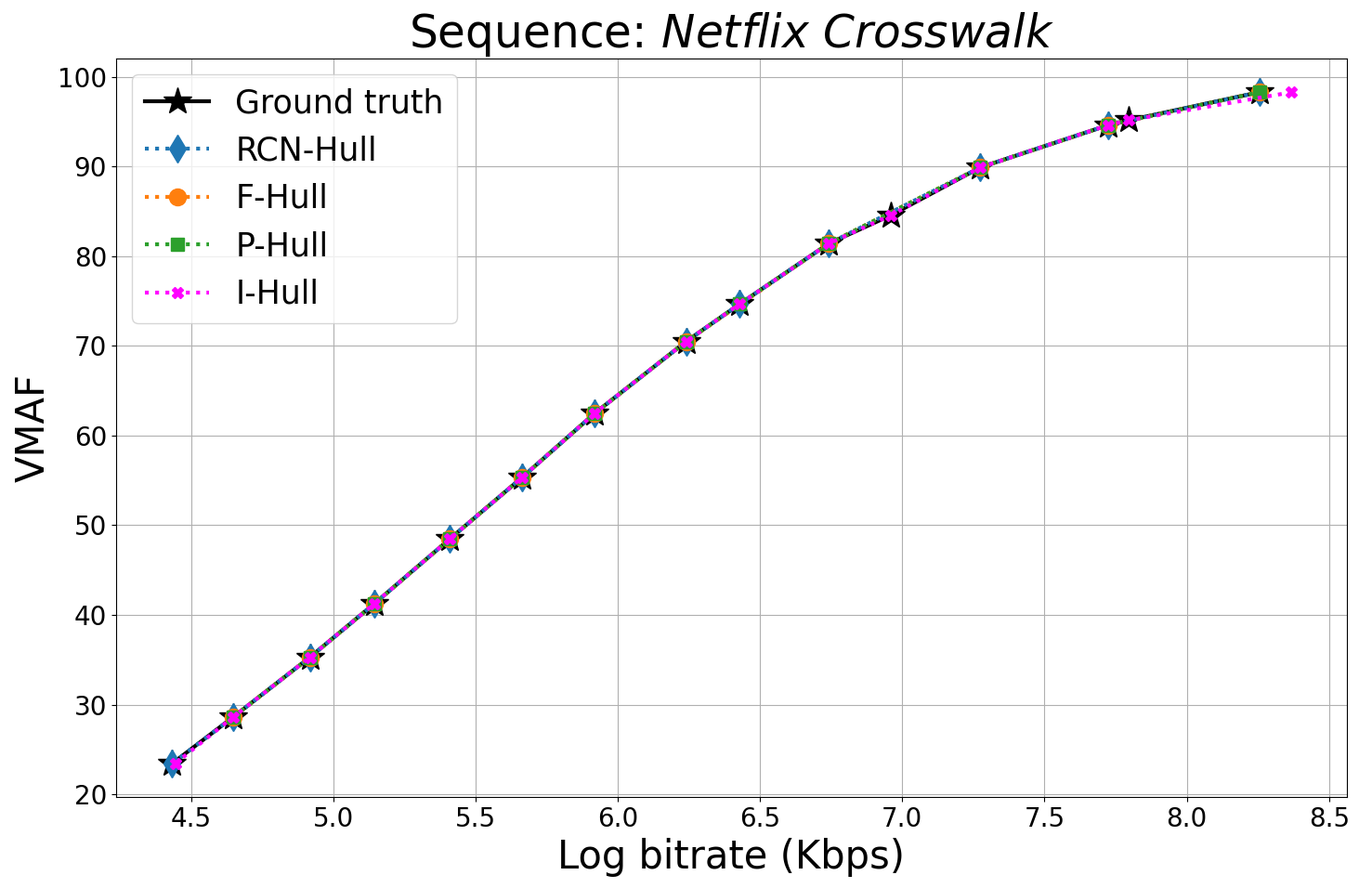}}
\hfil
\subfloat
    {\includegraphics[width=0.32\textwidth]{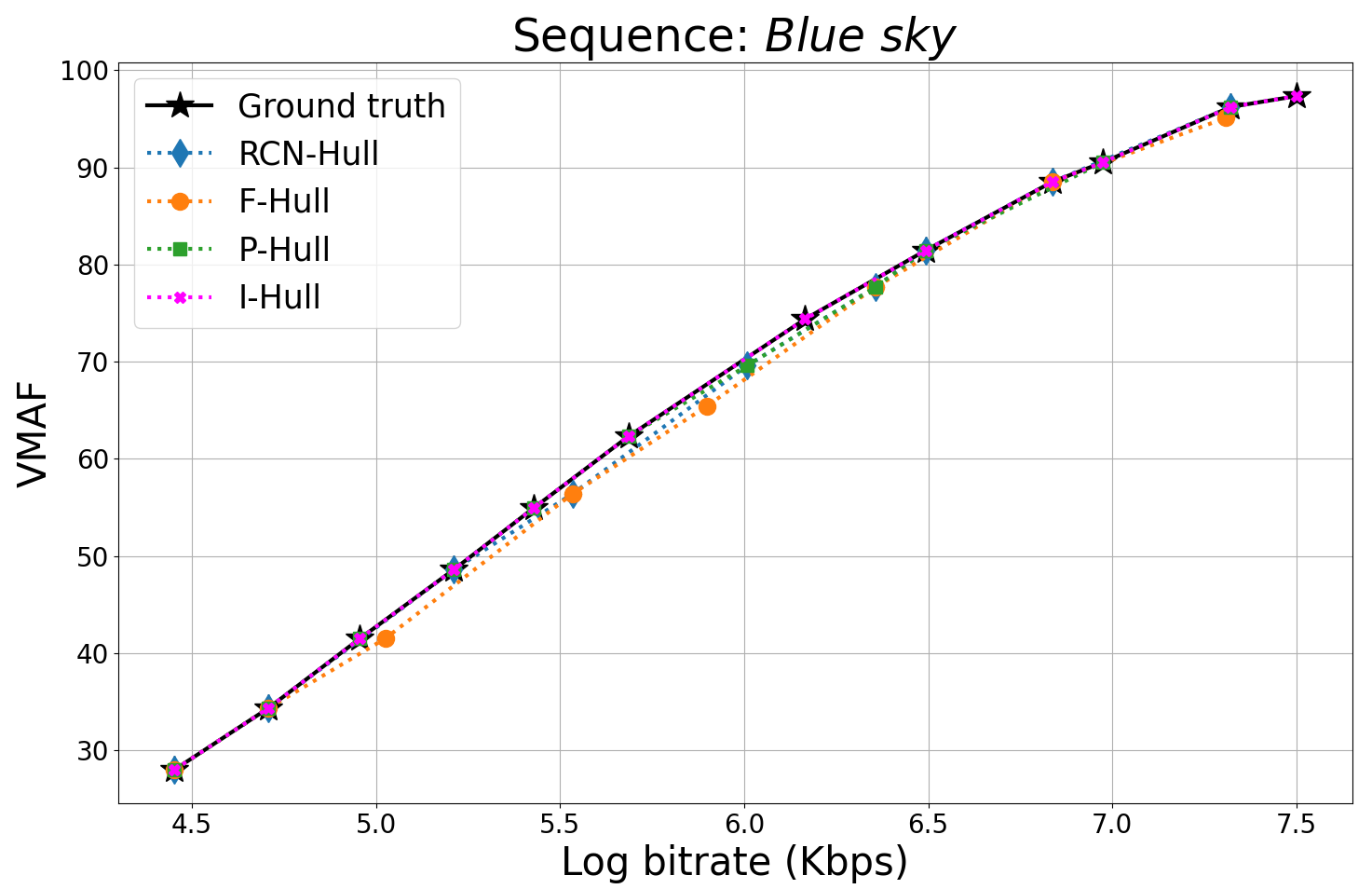}}     

    \caption{\small{{Plots of predicted bitrate ladders obtained using I-Hull, P-Hull \cite{proxy}, F-Hull \cite{spatiotemporal} and RCN-Hull, along with the corresponding ground truth convex hulls.}}}
   \label{fig:hull_plots}
\end{figure*}

\subsection{Performance Comparison}
We compared the performance of RCN-Hull against that of I-hull, P-hull \cite{proxy} and F-hull \cite{spatiotemporal}. The distribution of the BD-rates of each compared model on the UCV test set is plotted in Fig. \ref{fig:bdrate}. The box plots\footnote{\label{note1}the box boundaries represent the lower and upper quartiles of the corresponding data, while the whiskers extend to their minimum and maximum values} in Fig. \ref{fig:bdrate} show that  the BD-rate distribution obtained by the interpolation based method and RCN-Hull is close to the zero BD-rate mark, and also has low dispersions. The BD-rate distributions corresponding to the feature based method described in \cite{spatiotemporal} and the the proxy based method of \cite{proxy} deviate significantly from zero and has higher  dispersions. 

We report the average total times taken by each method to predict the convex hull points, and the corresponding total encode times separately, as shown in Fig. \ref{fig:time_barplot}, where the prediction time of each method is defined as:
\begin{itemize}
\item \textit{I-Hull}: the time taken to infer the bitrate and quality values for the encodes corresponding to the QP values in the set $\mathcal{Q} \setminus \mathcal{Q}^\prime$  using interpolation. 
\item \textit{P-Hull}: the time taken to encode the candidate set of points using the x265 encoder. 
\item \textit{F-Hull}: the sum of feature extraction time and the Gaussian process classifier's inference time. 
\item \textit{RCN-Hull}: the inference time of RCN-Hull model. 
\end{itemize}

Fig. \ref{fig:time_barplot} reveals that the prediction times of F-Hull and RCN-Hull were comparable. The prediction times of P-Hull were higher due to the x265 encoding that needed to be performed, while I-Hull had the lowest prediction times, since the interpolation operation itself is very fast. However, its total encoding time was significantly higher than the other methods, since it does not completely eliminate the pre-encoding step, as discussed later in this section. However, for all the methods, the prediction times were insignificant as compared to the total encoding times. 

The box plots the obtained reductions in encoding time required to generate bitrate ladders on the UCV test shown in Fig. \ref{fig:speed} reveal that the reductions achieved by P-Hull \cite{proxy}, F-Hull \cite{spatiotemporal} and RCN-Hull are quite comparable. This is because the complexities of each of these three methods are negligible as compared to the overall encoding time. In Section \ref{subsec:complexity}, we reported that the inference time of the RCN-Hull model is a very small fraction of the total encoding time. Similarly, feature extraction followed by prediction in  F-Hull \cite{spatiotemporal}, and the x265 encoding at \textit{medium} preset in P-Hull \cite{proxy} also consumes a very small amount of time as compared to the total encoding time. As a result, the differences in the encode time reductions achieved by these three methods arise primarily  from differences in their predictions. For example, a method that tends to predict fewer points, especially at higher resolutions as well as lower QPs, will result in greater time savings than models that predict more such points. 

Although the interpolation based method achieves the lowest dispersion in terms of BD-rates, the time savings attained by this scheme are much smaller than the other three methods, because a minimum of $|\mathcal{P}| \cdot |\mathcal{Q}^\prime| = 35$ pre-encodes are needed. Further, additional encodes are required corresponding to points  $(p,q)$ which lie on the convex hull and $\text{where} \ p \in \mathcal{P} \ \text{and} \ q \in \mathcal{Q} \setminus \mathcal{Q}^\prime$. Thus, the average number of encodes required by this scheme is much higher, which explains its limited speedup.

The overall performances of the four compared methods in terms of average BD-rates, average BD-rate magnitudes and time savings, along with their 95\% bootstrap CIs are summarized in Table \ref{table:summary}. Table \ref{table:summary} also reports the mean absolute deviations (MAD) and the standard deviations (SD) of BD-rates obtained for the four compared methods. The predictions produced by the RCN-Hull model approximate the optimal convex hull more closely and reliably than those delivered by the proxy based and feature based approaches as implied by the lower average BD-rate magnitudes and lower BD-rate dispersion values achieved in terms of both MAD and SD. The time savings achieved RCN-Hull, P-Hull and F-Hull were similar, significantly outperforming the interpolation based approach, as shown in the first row of Table  \ref{table:summary}. This suggests that our approach is able to obtain both accurate and fast approximations of the optimal convex hulls.  

The bitrate ladders generated using RCN-Hull on six test sequences is graphically illustrated in Fig. \ref{fig:hull_plots}\hspace{0.1cm}\footnote{\hspace{0.1cm}Higher resolution plots can be viewed at \url{https://drive.google.com/drive/folders/1wiCg3AoS0r3Les60yj8ZwrliVgG3OYVs?usp=sharing}.}, along with those predicted by I-Hull, P-Hull and F-Hull as well as the corresponding ground truth optimal convex hulls. Using Fig. \ref{fig:hull_plots}, it may be observed the curves produced by P-Hull and F-Hull contain higher percetanges of instances of predicted points not matching the ground truth, and more missed points, as compared to the  RCN-Hull predictions. 

\section{Conclusion}
\label{sec:conclusion}

We introduced a RCN based method of constructing content-aware bitrate ladders for adaptive streaming applications. We compiled a large collection of video shots and created a database of corresponding convex hulls generated by performing HEVC encodes at seven resolutions and nine QP values. A transfer learning scheme was employed in order to alleviate the limited availability of uncompressed video contents to train deep models. A net time savings of 54.6\% was achieved on the videos tested using the new RCN-Hull model, and the experimental analysis shows that predictions produced by it are closer to the theoretically optimal convex hulls as compared to contemporary approaches that use proxy encoders, and ML using handcrafted features. 

A direction of further research that is worthy of exploration is to apply RCN-Hull on encoding parameter spaces having additional degrees of freedom, such as the temporal sampling rate. Further, bitrate ladders comprising higher resolution videos, which are likely to dominate future trends in video streaming are also likely to be benefited by the use of RCN-Hull.

{\appendices

\section{Role of RCN in Model Performance}
Neural network architectures such as \cite{3dcnn}, \cite{C3D}, and \cite{I3D}, which process spatiotemporal video data using receptive fields of finite size, are not suitable for processing video shots having arbitrarily large numbers of high resolution frames as a single input, due to memory constraints. Thus, recurrent modeling was imperative to handle our use case of spatiotemporal complexity analysis of full video shots for convex hull prediction. However, to investigate the specific role of RCNs in determining model performance, we explored an alternative model architecture, where we decoupled the recurrence and convolutional properties of the model. One such architecture where the convolutional blocks and recurrent blocks are implemented as separate layers was introduced in \cite{rcn}, where visual features were first extracted by using a CNN block, followed by sequentially modeling the visual features using an RNN block. Adopting the same approach, we designed an alternative model architecture, which performed CNN based feature extraction, followed by sequential modeling of the resulting feature maps with GRUs. This architecture thus decoupled the spatial modeling task from the temporal modeling task, by contrast with the joint spatiotemporal modeling performed by the Conv-GRU units in our proposed RCN-Hull model. 

The number of trainable parameters of the decoupled CNN+RNN model is 554,779, which is commensurate with the complexity of our proposed RCN-Hull model. It was trained in the same way as described in Section \ref{sec:results}. Its prediction performance is reported in Table \ref{table:model_comp}, relative to that of the RCN-Hull model. 
\begin{table}[htb]
\caption{Classification performance metrics of the CNN+RNN model compared with RCN-Hull, along with 95\% CIs.}
\centering
\vspace{0.1cm}
 \scalebox{0.88}{
 \begin{tabular}{|c | c | c | c |} 
 \hline
  \multirow{2}{1cm}{\textbf{Approach}} &  \multirow{1}{1cm}{\textbf{Precision}}
  & \multirow{1}{1cm}{\hfil \textbf{Recall}} & \multirow{1}{1.1cm}{\hfil \textbf{F1 score}} \\
& \textbf{(\%)} & \textbf{(\%)} & \textbf{(\%)} \\
\hline
 \multirow{2}{1.5cm}{CNN+RNN} & 79.6  & 59.0 &  67.8 \\ 
 & [78.5, 80.7] & [57.6. 60.3] & [66.5, 69.0] \\
 \hline
\multirow{2}{1.5cm}{RCN-Hull}  & \textbf{91.6}  & \textbf{77.8} & \textbf{84.1}   \\ 
& [90.6. 91.9] & [76.0. 77.9] & [82.7, 84.3] \\
\hline
\end{tabular}
}
\label{table:model_comp}
\end{table}

Table \ref{table:model_comp} reveals that the prediction performance suffered considerably when the spatial and temporal analysis steps were decoupled, as in the CNN+RNN model. This experiment attests to the advantage of fusing the convolutional and recurrent modeling into a single unit, as done in the Conv-GRU RCN units, which comprise the building blocks of our RCN-Hull model. 

\section{RCN-Hull Performance on Short Video Shots}
\label{app: appb}

Since the RCN-Hull model was trained using 300 frames from each video shot, it is interesting to examine its performance on shorter video chunks. Thus, we evaluated the performance of the model on several 2-3s long segments, starting from the first frame of the corresponding sequences, as summarized in Table \ref{table:shorter}. The results presented in Table \ref{table:shorter} show that our model was able to make predictions that delivered small BD-rates and significant time savings, thereby further enforcing its consistent performance across the different lengths of video shots. 
\begin{table}[htb]
\caption{BD-rates and time savings of RCN-Hull test video shots of lengths 2-3s.}
\centering
\vspace{0.1cm}
 \scalebox{0.85}{
 \begin{tabular}{|c | c | c | c |c|} 
 \hline
  {\textbf{Sequence}} &  {\textbf{Duration}} &  {\textbf{\# of frames}}
  & {\hfil \textbf{BD-rate}} & \textbf{Time savings} \\
   
& (s)  &  &  \textbf{(\%)} &  \textbf{(\%)} \\
\hline
 \textit{Motor racing}  & 2  & 48 & -0.20 & 55.6 \\ 
 \textit{Basketball}  & 2  & 50 & 1.02 & 54.7  \\ 
   \textit{Dolphins}  & 2  & 50 & 0.0 & 47.8  \\ 
 \textit{Aspen}  & 2  & 60 & -0.02 & 58.6  \\ 

  \textit{Scarf}  & 2.5  & 75 & 0.77 & 57.7 \\ 
 \textit{Birds in cage}  & 2.5  & 75 & 0.50 & 57.6 \\ 
  \textit{Outdoor mall}  & 2.5  & 75 & 0.03 & 54.4 \\ 

  \textit{Wildlife}  & 3  & 75 & 0.40 & 54.3 \\ 
  \textit{Seeking}  & 3  & 75 & 0.0 & 58.5 \\ 
   \textit{Rush field cuts}  & 3  & 90 & 0.27 & 39.1 \\ 
   \hline
   \multicolumn{3}{|c|}{Average} & 0.28 & 53.8 \\ 
\hline
\end{tabular}
}
\label{table:shorter}
\end{table}

\section{Explanation for Negative Values of BD-rates}
\label{app: appc}

The convex hulls were constructed using the actual values of the bitrates in the linear scale. By the the convexity of the resulting ground truth optimal RQ curve (i.e. the convex hull), as well as due to the formulation of the convex hull prediction task as a multi-label classification problem, whereby we predict whether or not each of the 63 candidate RQ points lie on the convex hull, the RQ points predicted by the RCN-Hull model can either lie on or to the right of the ground truth were found to lie to the right of the corresponding ground truth RQ curve, when plotted using a linear bitrate scale, as demonstrated in Fig. \ref{fig:PCHIP_linear}. However, when bitrates are transformed to the logarithmic scale, the curves formed by interpolating the same set of RQ points that lie on the convex hull, are non-convex when plotted on this transformed bitrate scale, as illustrated in Fig. \ref{fig:PCHIP_log}. As a consequence of the non-convexity of the RQ curves arising due to the logarithmic transformation of the bitrate axis, when one or more RQ points that belong to the ground truth convex hull are excluded from the model predictions, the corresponding segment(s) of the predicted RQ curve can lie to the left of the ground truth RQ curve, as shown in the Fig. \ref{fig:PCHIP_log}. Since the calculation of BD-rates uses logarithmic values of bitrates, this effect manifests in the form of negative BD-rates that were reported in Tables \ref{table:performance} and \ref{table:summary} and in Fig. \ref{fig:comparison}. The curves shown in Fig.\ref{fig:PCHIP_linear} and \ref{fig:PCHIP_log} were obtained using PCHIP interpolation, which is also the interpolation method used to compute the BD-rates reported in the paper. The same effect was also observed when linear interpolation was used to construct the curves instead, as shown in Figs. \ref{fig:linear_linear} and \ref{fig:linear_log}. 

\begin{figure}[htb]
\centering
\subfloat[\label{fig:PCHIP_linear}]
    {\includegraphics[width=0.48\linewidth]{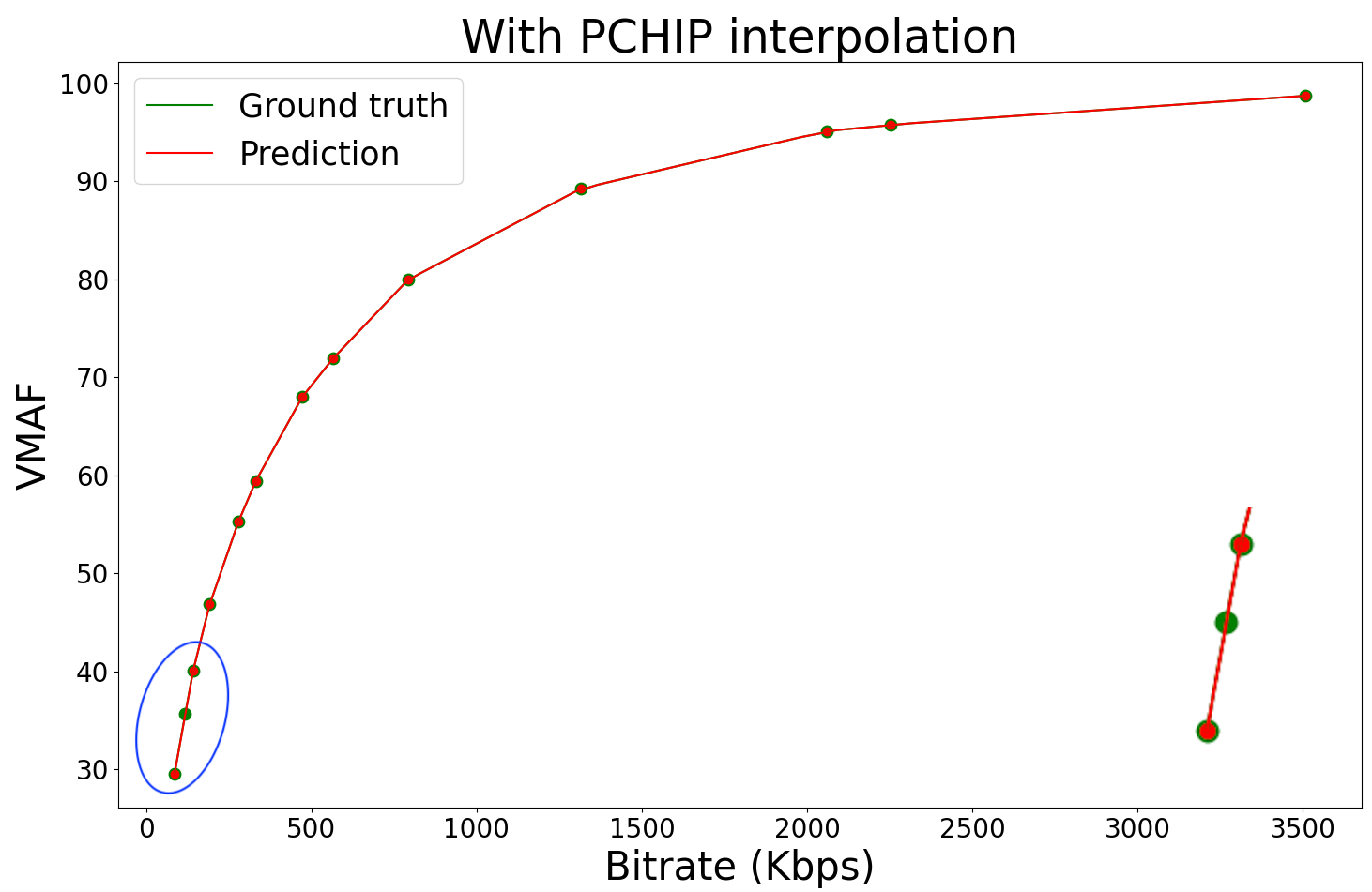}}
\subfloat[\label{fig:linear_linear}]
    {\includegraphics[width=0.48\linewidth]{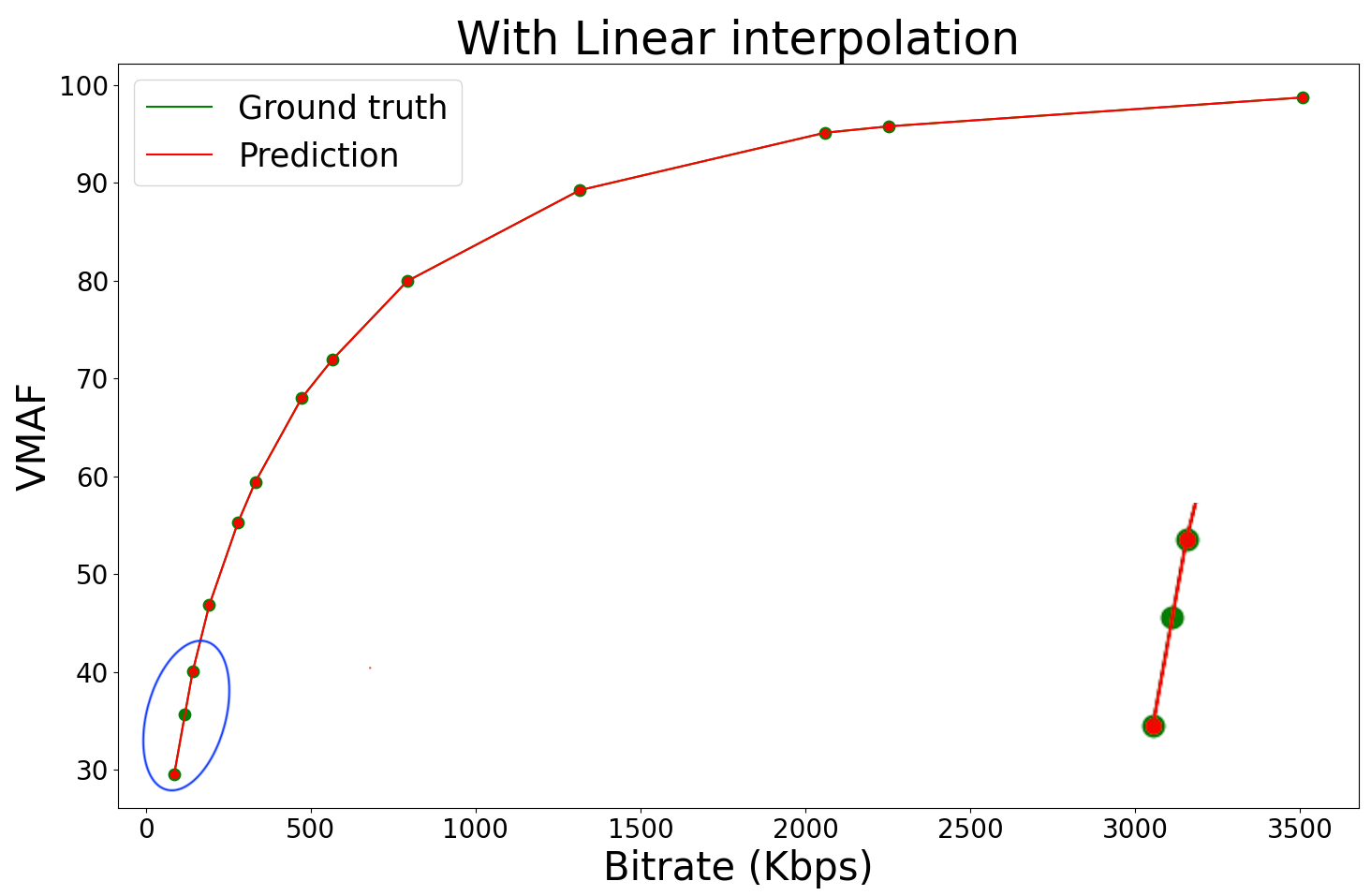}}
 \hfil
\subfloat[\label{fig:PCHIP_log}]
    {\includegraphics[width=0.48\linewidth]{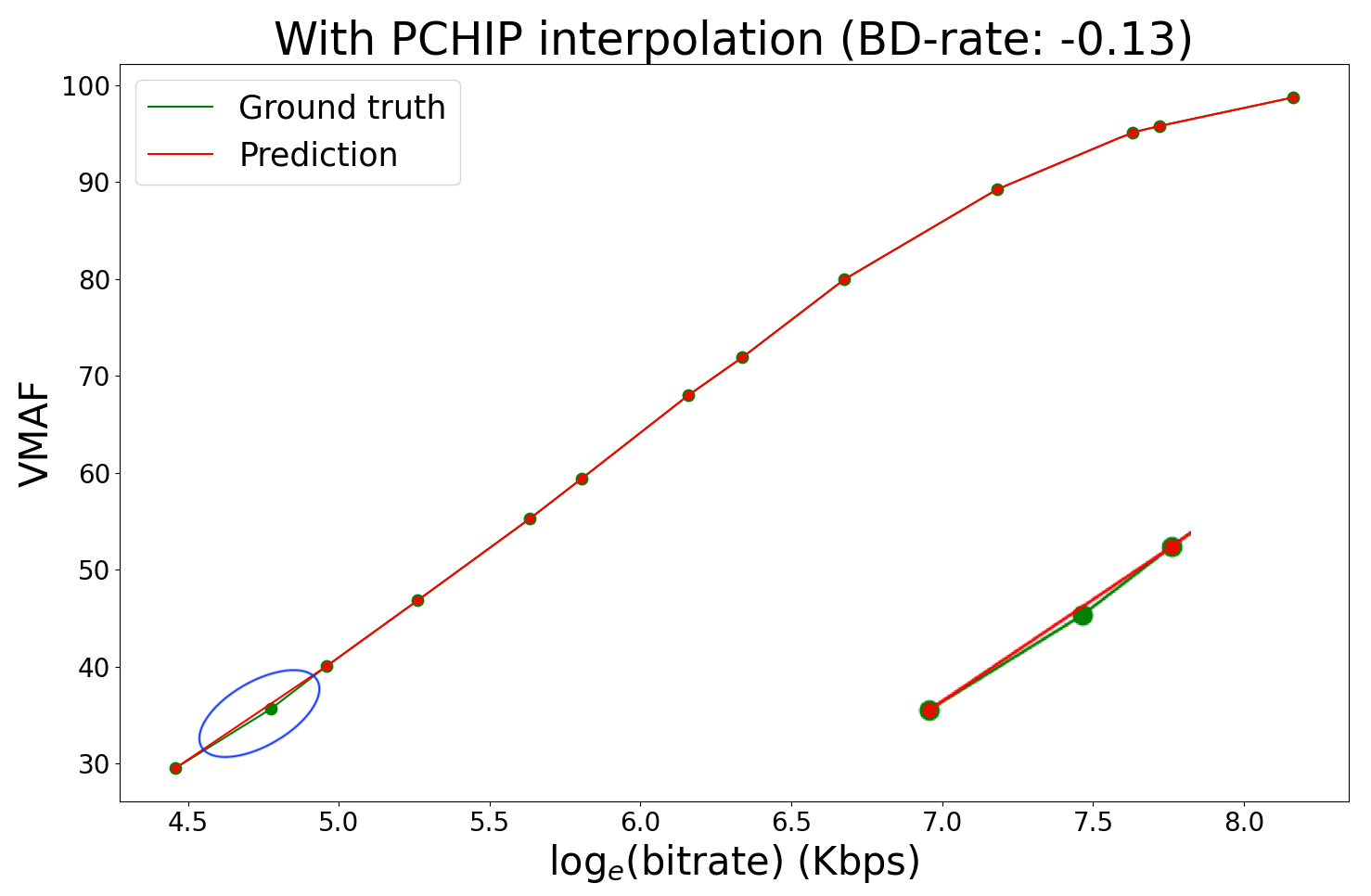}}
\subfloat[\label{fig:linear_log}]
    {\includegraphics[width=0.48\linewidth]{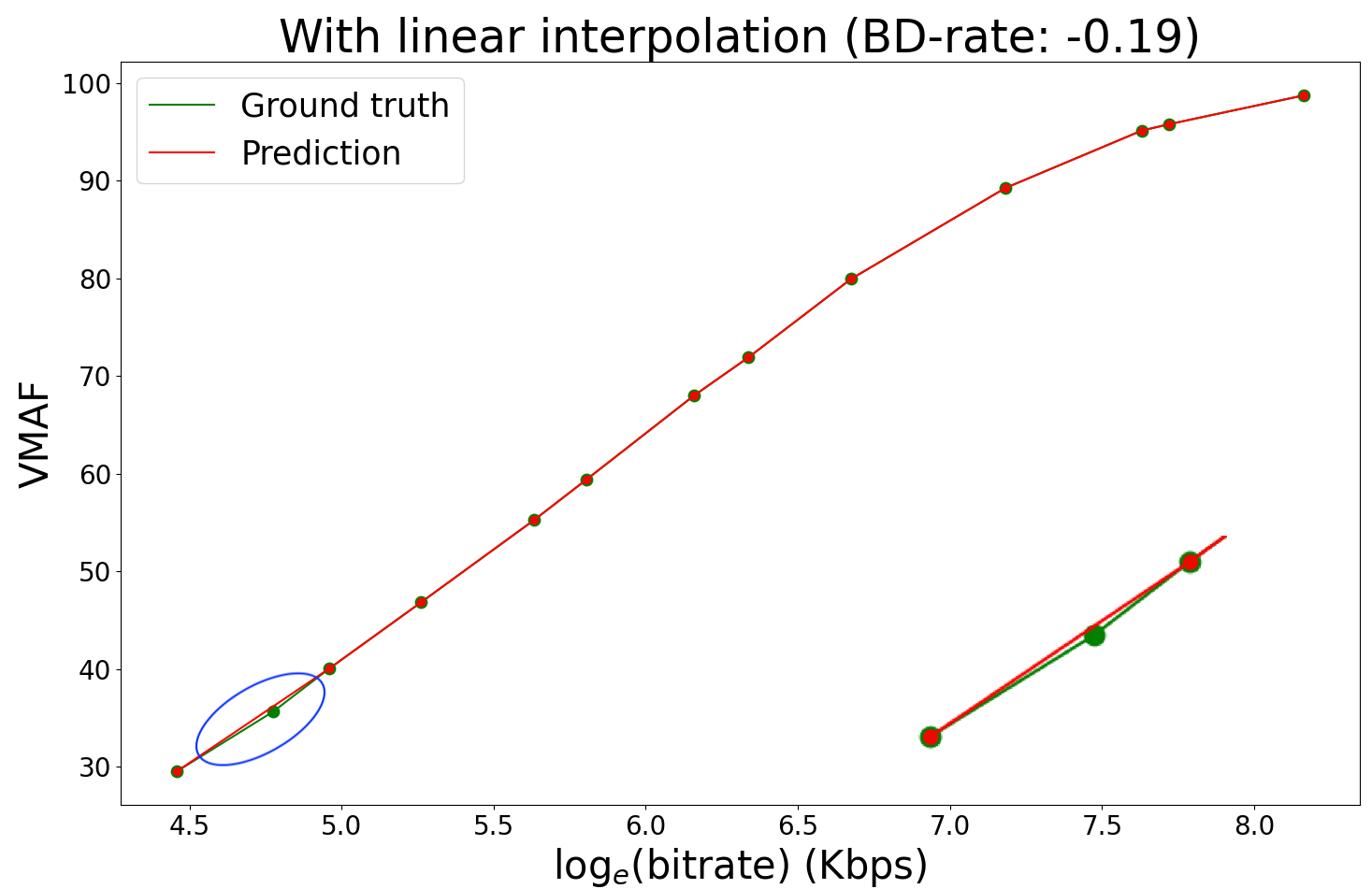}}

    \caption{\small{{Ground truth and predicted convex hulls plotted using linear ((a) and (b)) and logarithmic bitrate scales ((c) and (d)).}}}
   \label{fig:hulls_linlog}
\end{figure}

}  

%

\balance
\bibliographystyle{IEEEtran}
\bibliography{refs}

\end{document}